\documentclass[a4paper,twoside]{article}

\usepackage{epsfig}
\usepackage{subcaption}
\usepackage{calc}
\usepackage{amssymb}
\usepackage{amstext}
\usepackage{amsmath}
\usepackage{amsthm}
\usepackage{multicol}
\usepackage{pslatex}
\usepackage[round]{natbib}
\usepackage{algorithm2e}

\usepackage{mathptmx} 
\DeclareRobustCommand{\bigO}{%
  \ensuremath{\text{\usefont{OMS}{cmsy}{m}{n}O}}%
}

\usepackage{multicol}

\usepackage{csquotes}
\usepackage{tabularray}
\usepackage{listings}
\usepackage{tikz}
\usepackage{pgfplots}
\usetikzlibrary{arrows.meta, shapes.symbols, shapes.arrows, backgrounds, matrix, positioning, calc, pgfplots.groupplots}
\pgfplotsset{compat=1.18}
\usepackage[hidelinks]{hyperref}
\usepackage[capitalize, nameinlink]{cleveref}
\creflabelformat{equation}{#2\textup{#1}#3}
\crefname{section}{Sec.}{Secs.}

\usepackage{fancyhdr}
\newcommand{\quot}[1]{\emph{\enquote{#1}}}

\definecolor{fhgBlue}{HTML}{005C80}
\definecolor{fhgBlueLight}{HTML}{dfedf5}
\definecolor{fhgOrange}{HTML}{F58221}
\definecolor{fhgYellow}{HTML}{FDB627}
\definecolor{fhgRed}{HTML}{D33535}
\definecolor{fhgPurple}{HTML}{7C5ADC}
\definecolor{fhgPurpleLight}{HTML}{9D83E5}
\definecolor{fhg}{HTML}{179C7D}
\definecolor{fhgGreenLight}{HTML}{91CBBD}
\definecolor{pipeRoot}{HTML}{E2E2E2}

\definecolor{pipeReferenceA}{HTML}{f1dfd2}
\definecolor{pipeReferenceA1}{HTML}{e6a39f}
\definecolor{pipeReferenceA2}{HTML}{dfa88f}
\definecolor{pipeReferenceA3}{HTML}{d1ae84}
\definecolor{pipeReferenceA4}{HTML}{c0b581}
\definecolor{pipeReferenceB}{HTML}{d0e8dd}
\definecolor{pipeReferenceB1}{HTML}{93c094}
\definecolor{pipeReferenceB2}{HTML}{7fc2a6}
\definecolor{pipeReferenceB3}{HTML}{72c3ba}
\definecolor{pipeReferenceC}{HTML}{e0e1f5}
\definecolor{pipeReferenceC1}{HTML}{7ebee0}
\definecolor{pipeReferenceC2}{HTML}{94b8e7}
\definecolor{pipeReferenceC3}{HTML}{adb2e6}
\definecolor{pipeReferenceC4}{HTML}{c4abde}
\definecolor{pipeReferenceC5}{HTML}{d7a5d0}

\definecolor{pipeInitialB}{HTML}{efdfd1}
\definecolor{pipeInitialB1}{HTML}{e4a499}
\definecolor{pipeInitialB2}{HTML}{d4ad86}
\definecolor{pipeInitialB3}{HTML}{b9b781}
\definecolor{pipeInitialC}{HTML}{d4e5f4}
\definecolor{pipeInitialC1}{HTML}{75c3b3}
\definecolor{pipeInitialC2}{HTML}{6fc2d0}
\definecolor{pipeInitialC3}{HTML}{85bce3}
\definecolor{pipeInitialC4}{HTML}{a9b3e7}
\definecolor{pipeInitialC5}{HTML}{cba9da}

\definecolor{pipeShiftedB}{HTML}{dbe6d3}
\definecolor{pipeShiftedB1}{HTML}{beb581}
\definecolor{pipeShiftedB2}{HTML}{9fbd8b}
\definecolor{pipeShiftedB3}{HTML}{81c2a4}
\definecolor{pipeShiftedC}{HTML}{e9dff2}
\definecolor{pipeShiftedC1}{HTML}{7fbde1}
\definecolor{pipeShiftedC2}{HTML}{a2b5e8}
\definecolor{pipeShiftedC3}{HTML}{c6aadd}
\definecolor{pipeShiftedC4}{HTML}{dea3c6}
\definecolor{pipeShiftedC5}{HTML}{e7a2a8}

\definecolor{pipe3B}{HTML}{d7e7d6}
\definecolor{pipe3B1}{HTML}{b0b984}
\definecolor{pipe3B2}{HTML}{94c093}
\definecolor{pipe3B3}{HTML}{7bc3ab}
\definecolor{pipe3C}{HTML}{e6e0f3}
\definecolor{pipe3C1}{HTML}{7fbde1}
\definecolor{pipe3C2}{HTML}{9eb6e8}
\definecolor{pipe3C3}{HTML}{beade1}
\definecolor{pipe3C4}{HTML}{d8a5cf}
\definecolor{pipe3C5}{HTML}{e5a1b6}

\definecolor{pipe6B}{HTML}{d3e7d9}
\definecolor{pipe6B1}{HTML}{a2bd8a}
\definecolor{pipe6B2}{HTML}{89c19b}
\definecolor{pipe6B3}{HTML}{76c3b2}
\definecolor{pipe6C}{HTML}{e3e0f4}
\definecolor{pipe6C1}{HTML}{7fbee0}
\definecolor{pipe6C2}{HTML}{99b7e8}
\definecolor{pipe6C3}{HTML}{b6afe4}
\definecolor{pipe6C4}{HTML}{cfa8d7}
\definecolor{pipe6C5}{HTML}{e0a2c3}
\definecolor{interpReferenceA}{HTML}{f6dcda} 
\definecolor{interpReferenceA1}{HTML}{efbfc3} 
\definecolor{interpReferenceA2}{HTML}{edc1b7} 
\definecolor{interpReferenceA11}{HTML}{e7a2a8} 
\definecolor{interpReferenceA21}{HTML}{e5a39e} 
\definecolor{interpReferenceA22}{HTML}{e3a597} 

\definecolor{interpIncrementalA}{HTML}{cde8e5} 
\definecolor{interpIncrementalA1}{HTML}{d9c9a8} 
\definecolor{interpIncrementalA2}{HTML}{b9cdee} 
\definecolor{interpIncrementalA11}{HTML}{c7b281} 
\definecolor{interpIncrementalA21}{HTML}{6fc3c2} 
\definecolor{interpIncrementalA22}{HTML}{a3b4e8} 

\definecolor{interpNaiveA}{HTML}{e2e4d0} 
\definecolor{interpNaiveA1}{HTML}{e9c3b1} 
\definecolor{interpNaiveA2}{HTML}{e4c1dd} 
\definecolor{interpNaiveA11}{HTML}{dea88e} 
\definecolor{interpNaiveA21}{HTML}{aeba84} 
\definecolor{interpNaiveA22}{HTML}{dca3ca} 

\definecolor{interpOffsetA}{HTML}{e2e4d0} 
\definecolor{interpOffsetA1}{HTML}{e9c3b1} 
\definecolor{interpOffsetA2}{HTML}{b0d4b9} 
\definecolor{interpOffsetA11}{HTML}{dea88e} 
\definecolor{interpOffsetA21}{HTML}{aeba84} 
\definecolor{interpOffsetA22}{HTML}{85c29e} 

\newcommand{\tc}{\emph{Tree Colors}}
\newcommand{\cuttle}{\emph{Cuttlefish}}
\newcommand{\dtc}{\emph{DTC}}

\usepackage{SCITEPRESS}     

\begin{document}
\title{Dynamic Tree Colors: Adaptive Discriminable Hierarchies with Minimum Instability}

\author{\authorname{Tobias Mertz\sup{1,2}\orcidAuthor{0000-0001-5284-4350}, Steven Lamarr Reynolds\sup{1}\orcidAuthor{0000-0002-0564-2788} and J{\"o}rn Kohlhammer\sup{1,2}\orcidAuthor{0000-0003-1706-8979}}
\affiliation{\sup{1}Fraunhofer IGD, Darmstadt, Germany}
\affiliation{\sup{2}TU Darmstadt, Darmstadt, Germany}
\email{\{tobias.mertz, steven.lamarr.reynolds, joern.kohlhammer\}@igd.fraunhofer.de}
}

\keywords{Visual Analytics, Perception and Cognition in Visualization, Color Maps, Hierarchical Data.}

\abstract{Hierarchical color maps can support users in the analysis of hierarchical data.
For large hierarchies, dynamic color maps can improve discriminability upon user interactions, but the incremental color changes may cause users to lose their orientation in the data set.
To address this challenge, we present Dynamic Tree Colors, a dynamic hierarchical color map that can be configured to a suitable tradeoff between discriminability and color stability.
We also define quality metrics for both criteria and investigate our algorithm's performance with respect to these metrics as well as a user study with 18 participants.
Our results indicate that Dynamic Tree Colors yields good results in a wide range of application scenarios, but it does not achieve the performance of the state-of-the-art algorithm Cuttlefish in the specific scenario that algorithm was designed for.}

\onecolumn \maketitle \normalsize \setcounter{footnote}{0} \vfill
\thispagestyle{fancy}
\pagestyle{fancy}
\setlength{\headwidth}{\textwidth}

\section{\uppercase{Introduction}}
\label{sec:introduction}

\begin{figure*}
    \centering
    \scriptsize
    \setlength{\belowcaptionskip}{0pt}
    \begin{subfigure}{0.5\linewidth}
        \centering
        \begin{tikzpicture}
            \node[inner xsep=2pt, outer sep=0pt] (world) at (0,0) {\includegraphics[width=0.31\linewidth]{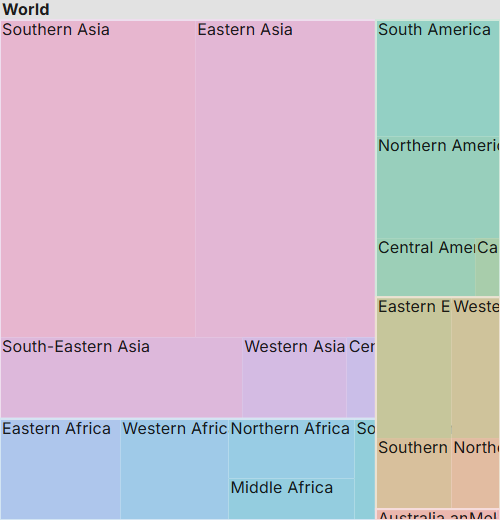}};
            \node[anchor=south] at (world.north) {Entire World};
            \node[inner xsep=2pt, outer sep=0pt, anchor=west] (americas) at (world.east) {\includegraphics[width=0.31\linewidth]{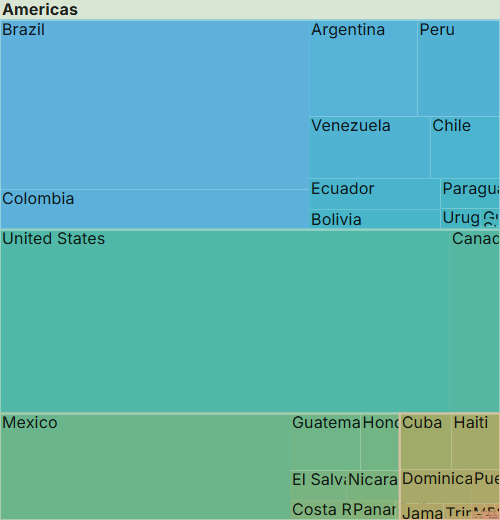}};
            \node[anchor=south] at (americas.north) {Region: Americas};
            \node[inner xsep=2pt, outer sep=0pt, anchor=west] (centralAmerica) at (americas.east) {\includegraphics[width=0.31\linewidth]{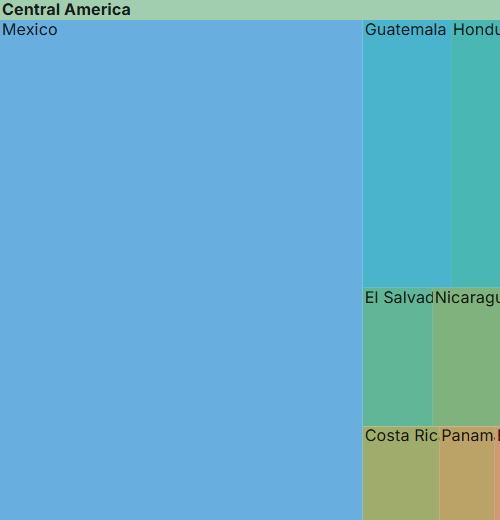}};
            \node[anchor=south] at (centralAmerica.north) {Sub-Region: Central America};
        \end{tikzpicture}
        \caption{Treemap (Stability Ratio: 0.4)}
        \label{fig:example:treemap}
    \end{subfigure}%
    \begin{subfigure}{0.5\linewidth}
        \centering
        \begin{tikzpicture}
            \node[draw=fhgBlueLight, inner xsep=1pt, outer ysep=3pt, outer xsep=1pt] (all) at (0,0) {\includegraphics[width=0.31\linewidth]{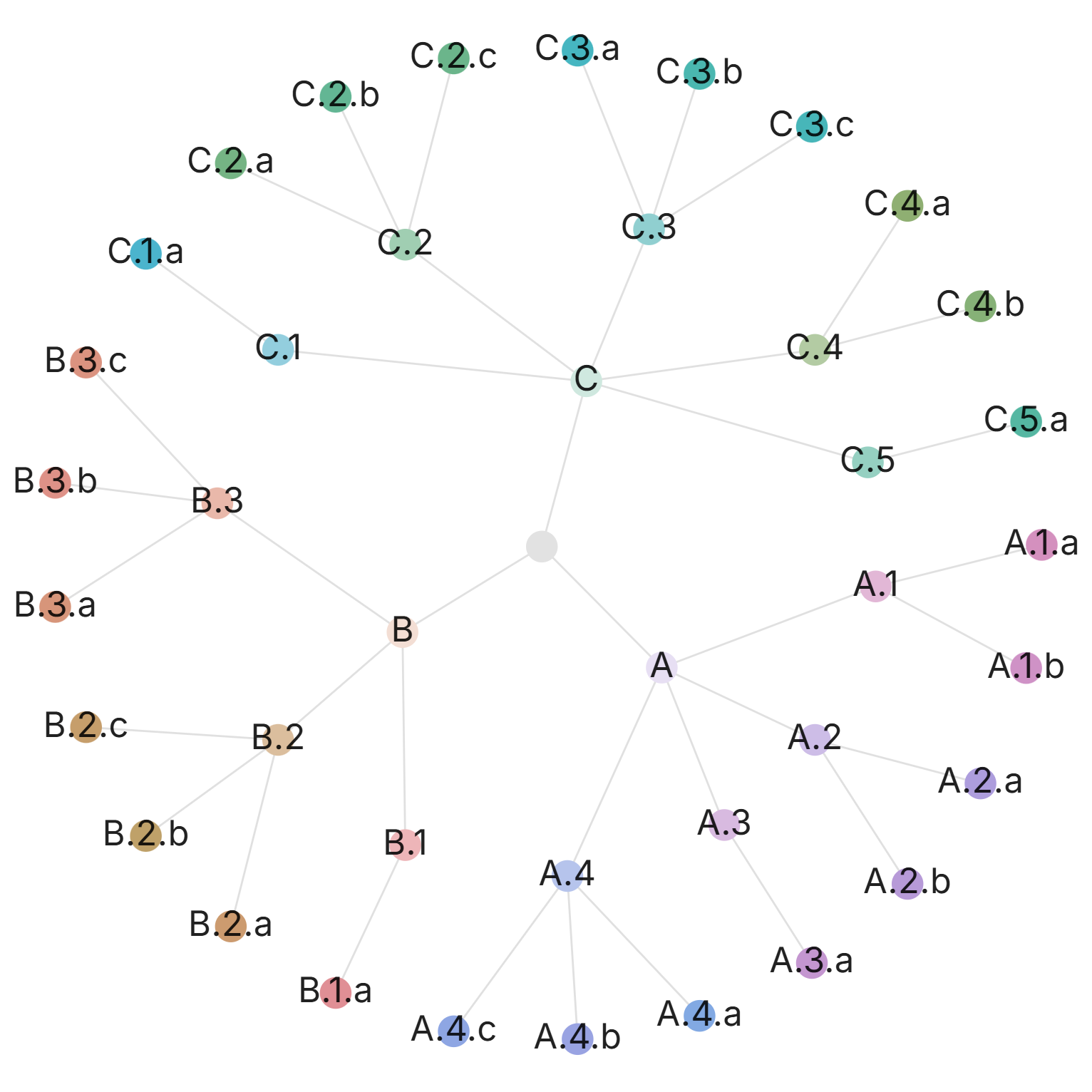}};
            \node[anchor=south] at (all.north) {Entire Hierarchy};
            \node[draw=fhgBlueLight, inner xsep=1pt, outer ysep=3pt, outer xsep=1pt, anchor=west] (noA) at (all.east) {\includegraphics[width=0.31\linewidth]{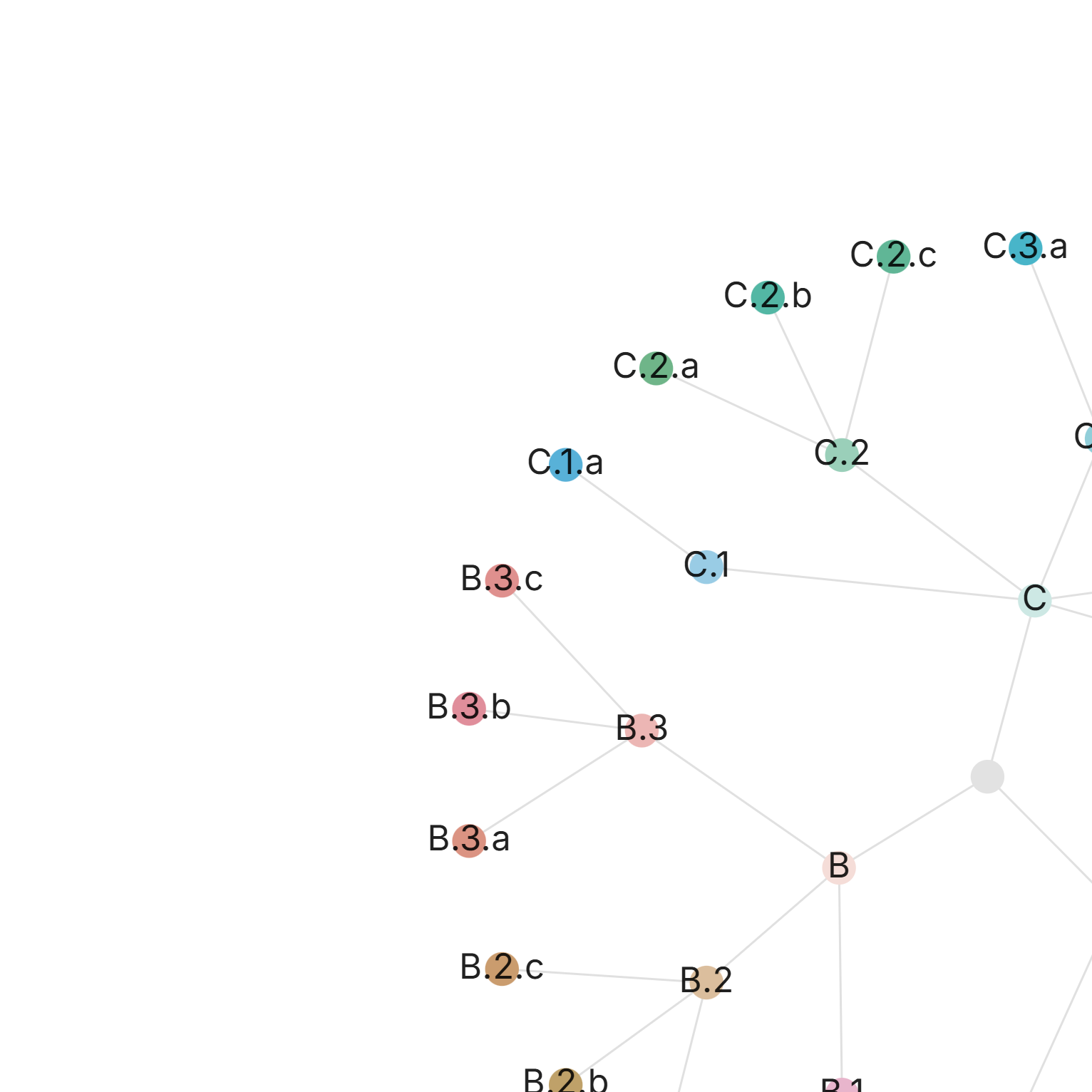}};
            \node[anchor=south] at (noA.north) {A Invisible};
            \node[draw=fhgBlueLight, inner xsep=1pt, outer ysep=3pt, outer xsep=1pt, anchor=west] (c) at (noA.east) {\includegraphics[width=0.31\linewidth]{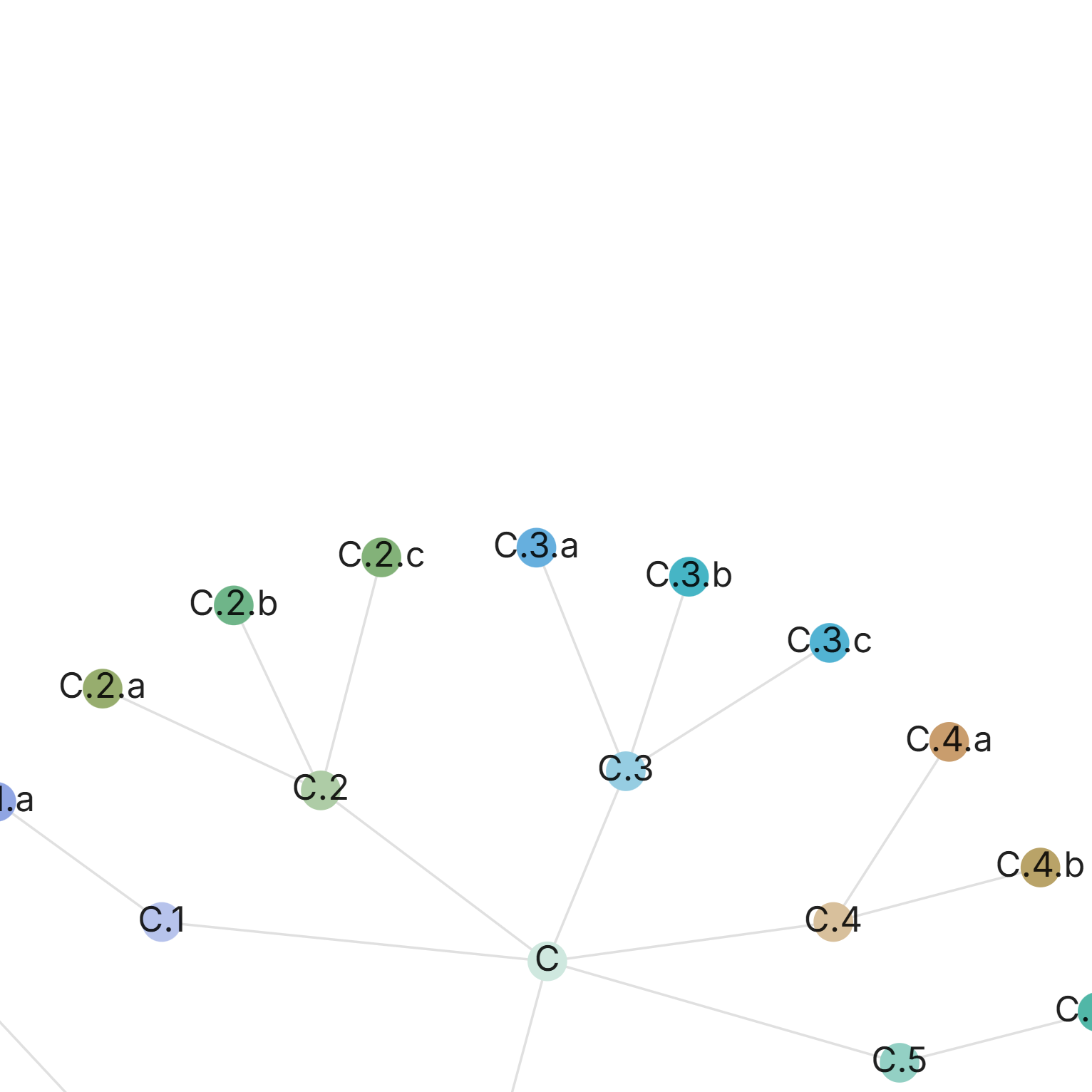}};
            \node[anchor=south] at (c.north) {C Focused};
        \end{tikzpicture}
        \caption{Node-Link Diagram (Stability Ratio: 0.5)}
        \label{fig:example:nodelink}
    \end{subfigure}
    \begin{subfigure}{\linewidth}
        \centering
        \begin{tikzpicture}
            \node[inner sep=0pt, outer sep=0pt] (all) at (0,0) {\includegraphics[width=0.33\linewidth]{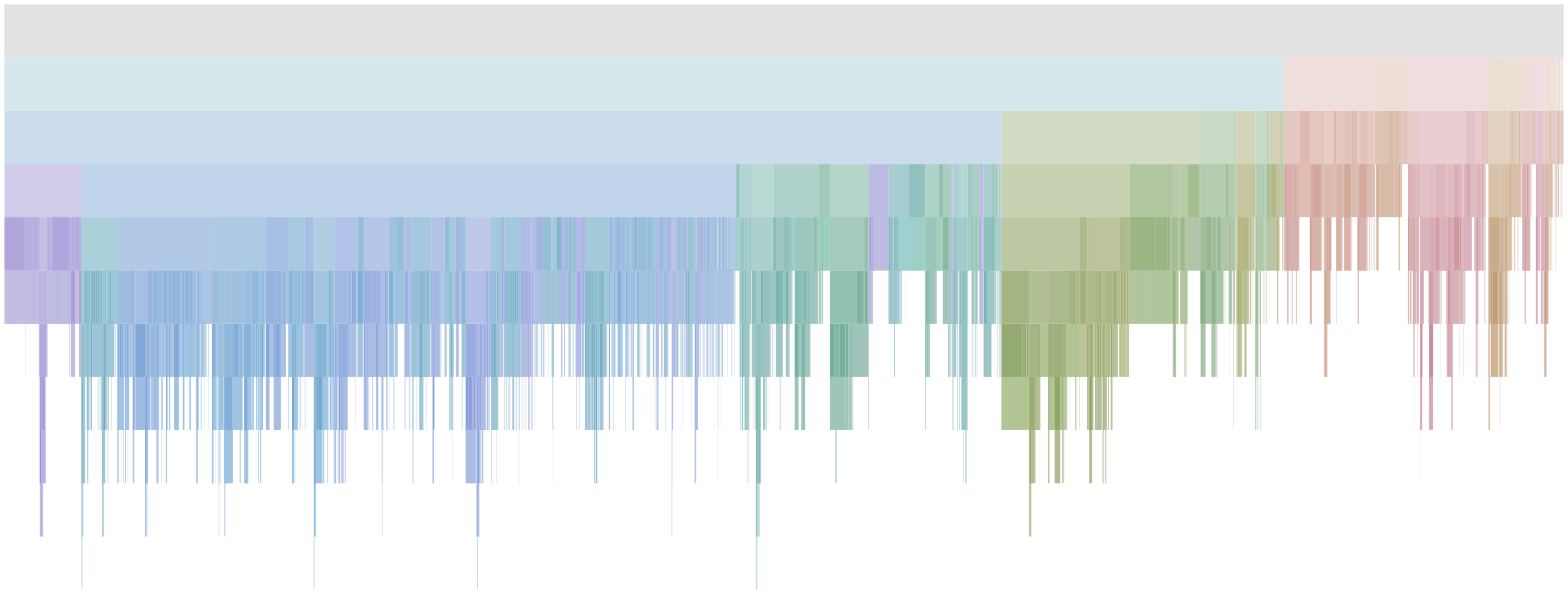}};
            \node[anchor=south] at (all.north) {Entire Hierarchy};
            \node[inner sep=0pt, outer sep=0pt, anchor=west] (com) at (all.east) {\includegraphics[width=0.33\linewidth]{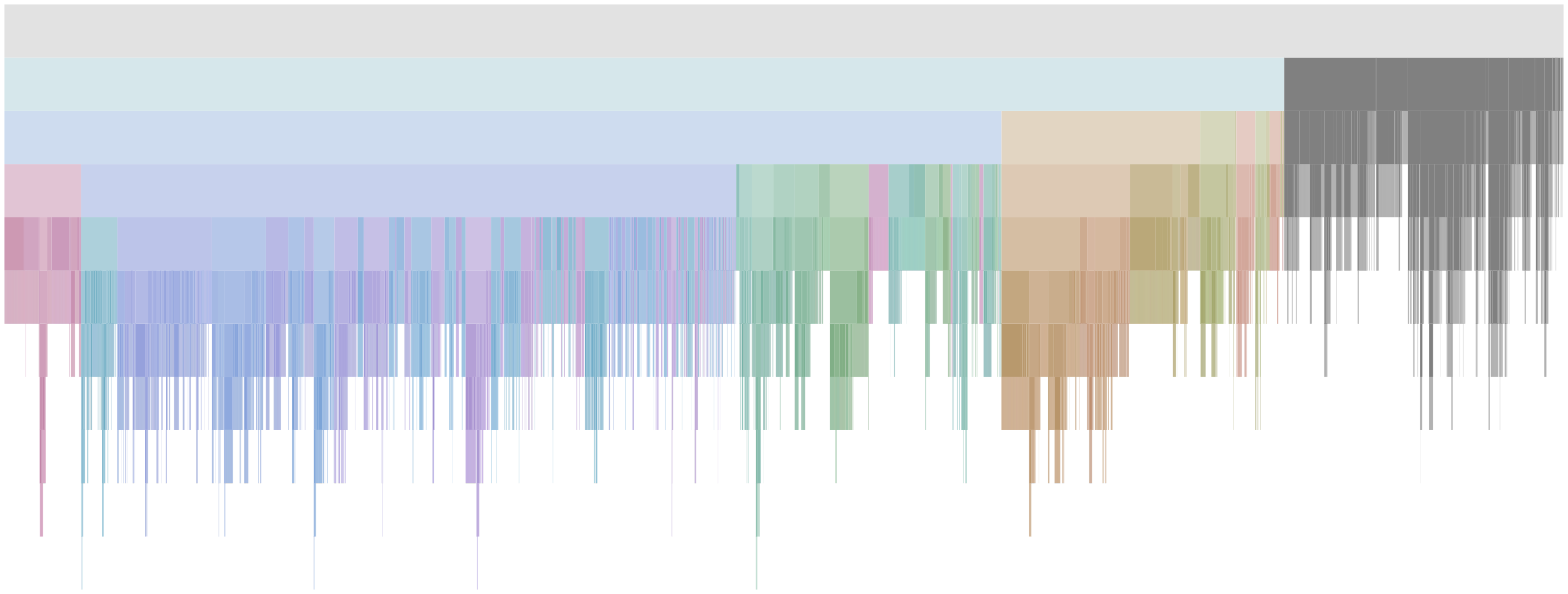}};
            \node[anchor=south] at (com.north) {.com Global Namespace Focused};
            \node[inner sep=0pt, outer sep=0pt, anchor=west] (amazon) at (com.east) {\includegraphics[width=0.33\linewidth]{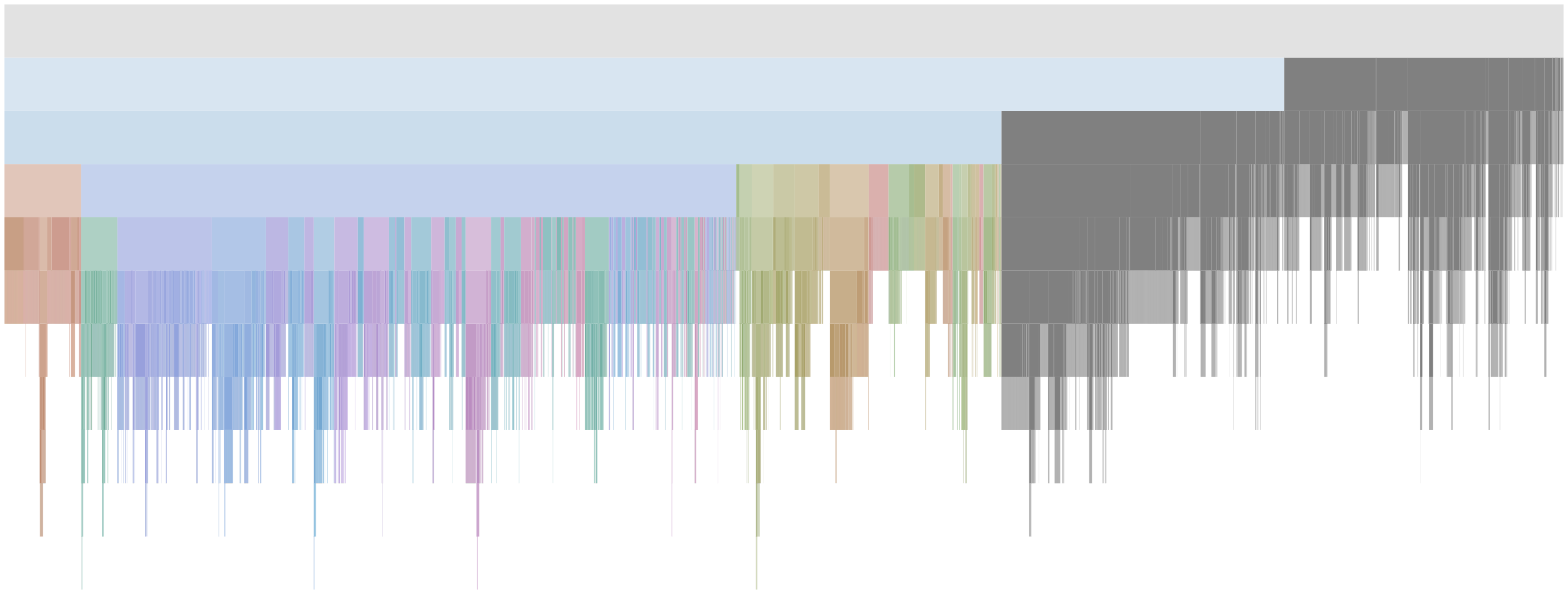}};
            \node[anchor=south] at (amazon.north) {Amazon Namespace Focused};
        \end{tikzpicture}
        \caption{Icicle Plot (Stability Ratio: 0)}
        \label{fig:example:icicle}
    \end{subfigure}%
    \setlength{\belowcaptionskip}{-4pt}%
    \caption{\dtc{ }applied to three application scenarios, each showing incrementally zoomed in interaction states from left to right. The stability ratio constrains the amount of color change between interaction states.}
    \label{fig:example}
\end{figure*}
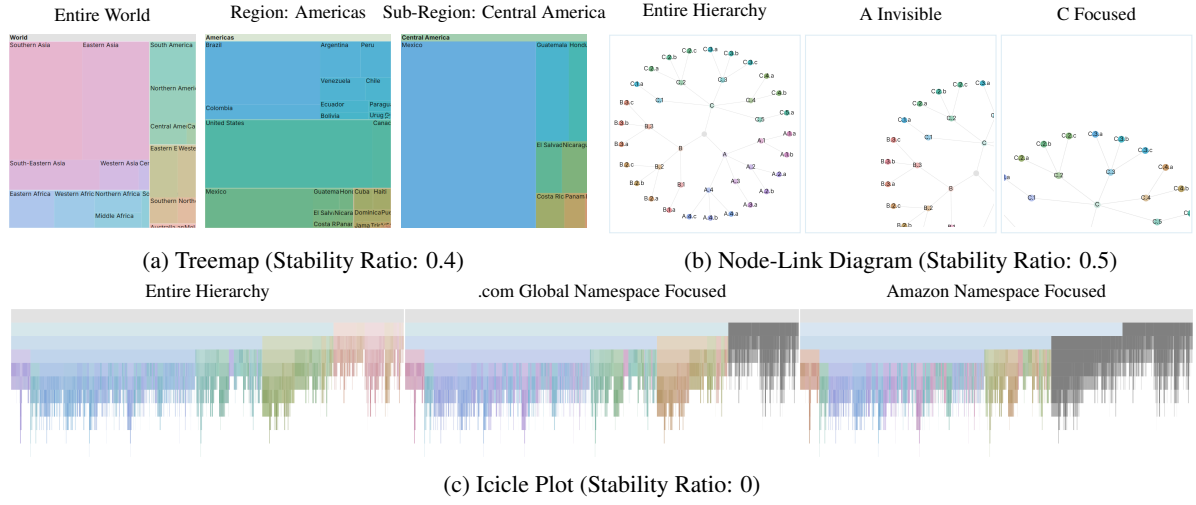

Hierarchical color maps can enhance the perception of hierarchical visualizations (see examples in \cref{fig:example}) or communicate hierarchical structures within non-hierarchical visualizations.
Over the years, several hierarchical color map generation algorithms have been proposed, the most popular of which is \tc~\citep{tennekesTreeColorsColor2014}.
This algorithm employs the nearly perceptually uniform HCL color space (Hue, Chroma, and Luminance) by recursively dividing the hue scale among siblings, assigning each node the center hue of its range.
Luminance and chroma vary linearly over the hierarchy levels.
\tc{ }achieves three design goals:
It assigns a unique color to each node, it encodes parent-child relations via similar colors, and it encodes the depth of a node in its color.
However, for large hierarchies, \tc{ }achieves poor discriminability.

To support larger hierarchies, dynamic algorithms adjust colors upon user interactions.
For example, the \cuttle{ }algorithm is designed for multi-scale visualizations~\citep{waldinCuttlefishColorMapping2019}, where the data is represented by a hierarchical structure of granularity levels.
Users can successively zoom into this hierarchy, thereby reducing the visible portion of the data set while increasing granularity (see \cref{fig:example:treemap}).
This allows the computation of a partial color map for the visible hierarchy nodes, which results in more discriminable colors.
To that end, \cuttle{ }assigns equidistant hues to the visible nodes of the currently visible granularity, with gaps between groups of siblings.
The angles between groups and nodes are computed based on the number of visible items on the hierarchy level and are constrained with a maximum threshold.
But the iterative adjustment of color maps introduces a design challenge:
The algorithm should keep node colors consistent between interaction states, to avoid disorienting users.
To achieve this, \cuttle{ }applies a rigid rotation to the hues, minimizing the hue differences between nodes and their parents, which were computed in an earlier interaction state.
This approach can improve color stability throughout user interactions, but it is limited to multi-scale visualizations, which only display one hierarchy level at a time.
As of today, there exists no such algorithm designed for general visualizations of hierarchical data.

We address this gap with \emph{Dynamic Tree Colors} (\dtc), a dynamic color mapping algorithm for visualizations of entire hierarchies.
The design goal of \dtc{ }is to combine the strengths of \tc{ }and \cuttle, which we tackle by combining the two approaches in a single algorithm.
For dynamic algorithms, the two main goals---increasing discriminability through dynamic color adjustment, and assuring color stability---are at odds.
Thus, practitioners need to find an appropriate tradeoff between the two criteria.
To discuss this tradeoff and provide empirical insight for color map designers, we introduce quantitative metrics for both quality criteria.
We analyze our approach in comparison with \cuttle{ }based on these metrics and perform a qualitative user study to investigate the relationship between our metrics and user perception.
Our main contributions are:

\begin{itemize}
    \item The \emph{Dynamic Tree Colors} algorithm.
    \item The definition of quantitative metrics for discriminative power and color stability in hierarchical color maps.
    \item A qualitative user study, comparing our algorithm with \cuttle{ }in diverse application scenarios.
\end{itemize}

\section{\uppercase{Related Work}}

\subsection{Hierarchical Color Maps}

In visualization research, the proximity-based coloring for multi-scale parallel coordinates is probably the earliest publication on hierarchical color maps~\citep{fuaHierarchicalParallelCoordinates1999}.
The algorithm assigns equidistant hues to nodes of a cluster hierarchy, via an in-order tree traversal.
Similarly, the structure-based coloring in InterRing~\citep{yangInterRingInteractiveTool2002} applies equidistant colors to the hierarchy leaves, while the inner node colors are computed from the weighted average of their children.
Conversely, the Hyperbolic Wheel~\citep{lamHyperbolicWheelNovel2012} inverts this approach by initializing colors for the direct children of the root, spaced according to their angular size in the radial visualization.
Descendant hues are then determined from their parent's hue with an additive offset and brightness is linearly decreased throughout the hierarchy levels.

These algorithms have been superseded by \tc~\citep{tennekesTreeColorsColor2014}, the state-of-the-art algorithm for static hierarchical color maps.
It combines the ideas of its predecessors to arrive at the recursive hue-subdivision approach described in \cref{sec:introduction}.
But static color maps suffer from poor discriminability for large hierarchies.

To tackle this challenge, the development of dynamic algorithms was spearheaded by \citet{waldinChameleonDynamicColor2016} with \emph{Chameleon}, which they later reworked to result in \cuttle~\citep{waldinCuttlefishColorMapping2019}.
\cuttle{ }replaces the force-based hue assignment of \emph{Chameleon} with the more efficient rigid rotation approach described in \cref{sec:introduction}.
Recently, \citet{chenDynamicColorAssignment2025} proposed to move to a linear optimization approach, to incorporate color harmony as second optimization objective.
But all three dynamic approaches are designed for multi-scale visualizations that only display one hierarchy level at a time, instead of visualizations for entire hierarchies.
Our approach combines the strengths of \tc{ }with a dynamic approach to create discriminable colors for the entire hierarchy.

\subsection{Color Map Quality}

Many publications have proposed design rules and quality criteria for color maps in the past.
For quantitative color maps, this includes surveys that collect the vast amount of research concerning one dimensional~\citep{bujackGoodBadUgly2018} or two dimensional~\citep{bernardSurveyTaskbasedQuality2015} color maps.
In contrast, categorical color maps have received less attention, with the underlying principle remaining: keep chroma and luminance constant and vary hue across the categories~\citep{harrowerColorBrewerorgOnlineTool2003, zeileisEscapingRGBlandSelecting2009, bartolomeoReflectionsUsesAvailable2025}.
Based on this approach, many additional criteria have been proposed, such as mark type~\citep{szafirModelingColorDifference2018}, color concept associations~\citep{rathoreEstimatingColorConceptAssociations2020}, color names~\citep{luPalettailorDiscriminableColorization2021}, object proximity in the visualization~\citep{liColorAssignmentOptimization2023}, and aesthetic criteria~\citep{gramazioColorgoricalCreatingDiscriminable2017}.

For hierarchical color maps, the research is very limited.
In a previous work, we have translated some of the most common design rules from the literature into terms of hierarchical color maps, but we did not define quantitative metrics~\citep{mertzQualityApproachHierarchical2024}.
Here, we build upon these design rules and the proposed metrics by \citet{bujackGoodBadUgly2018}, to define metrics for hierarchical color maps.
We focus on the discriminative power, which is generally agreed upon to be the most important quality criterion~\citep{chenDynamicColorAssignment2025}, and the color stability, which is the defining challenge for dynamic color map algorithms.

\input{pipeline.tex}

\section{\uppercase{The Algorithm}}
\label{sec:algorithm}

\Cref{fig:pipeline} shows the pipeline of the \dtc{ }approach.
We incrementally adjust the hierarchy colors simply by re-applying the \tc{ }algorithm upon every user interaction (\cref{fig:pipeline:initial}).
To ensure color stability, we align the resulting colors with a static reference color map (\cref{fig:pipeline:align}), which we compute once during initialization (\cref{fig:pipeline:reference}) and which includes unique colors for every node in the hierarchy.
Finally, to achieve the best tradeoff between discriminative power and color stability, we interpolate between the incremental and static reference colors (\cref{fig:pipeline:interpolate}).
In the following, we will describe the individual steps in detail.

\subsection{Computing Initial Colors}

The pipeline has two entry points, where the static \tc{ }algorithm is used to initialize the computation.
The first entry point is triggered upon initial start-up.
Here, we compute the static reference color for each node in the hierarchy (\cref{fig:pipeline:reference}).
Upon user interaction, the second entry point triggers the computation of the incremental colors for the currently visible portion of the hierarchy (\cref{fig:pipeline:initial}).
Here, the algorithm can allot a greater range of hues to the visible nodes than in the reference color map, because the invisible nodes do not occupy any color space.
This improves discriminability once users reduce the amount of visible nodes.
Our algorithm is compatible with arbitrary configurations of the \tc{ }approach, but the reference and incremental colors must be computed with the same parameters.
When applying \tc{ }with permutations and reversals of sibling node colors~\citep{tennekesTreeColorsColor2014}, we must make sure that the ordering of node-colors remains consistent with the reference color map.
To this end, we always apply the reference's sibling order.

\subsection{Applying Color Rotation}

In the first refinement step (\cref{fig:pipeline:align}), we align the incremental colors with the reference colors to reduce hue differences.
To that end, we employ the same concept as \cuttle.
We rotate the colors in HCL by the weighted average difference between each node's incremental and reference hue value:

\begin{equation}
    \gamma = \frac{\sum_{i \in S} ((h_R(v_i) - h_S(v_i)) \cdot \omega_i)}{\sum_{i \in S} \omega_i}\label{eq:diffAngle}
\end{equation}

Here, $h_R$ and $h_S$ are hue-assignment functions for the reference ($R$) and the current interaction state ($S$) for a given hierarchy node (vertex) $v$.
$\omega$ is a weighting factor, which can be adjusted to put emphasis on the stability of certain nodes.
An interaction state is defined as the set of visible nodes, thus $S \subseteq R$.
As the hue axis is circular, we must make sure to always use the smaller of the two possible distance values in this computation.
In contrast to \cuttle, we compute $\gamma$ considering all visible nodes in the hierarchy, not just those on an individual hierarchy level, and we compute the difference between incremental and reference hues as opposed to the difference between child and parent hue.
We apply the rotation by adding $\gamma$ to each node's incremental hue value.
This is a rigid transformation of the node colors, which means that it can improve color stability without negatively impacting the inter-node distances and, thus, discriminability.

\subsection{Configuring the Tradeoff}

The final step of the algorithm (\cref{fig:pipeline:interpolate}) interpolates between the reference and incremental colors to achieve a configurable tradeoff between color stability and discriminative power.
To that end, we view the incremental and reference colors as two extremes in a continuous spectrum, which we call the \textbf{stability ratio}.
We define the reference colors to correspond to a stability ratio of $1$, resulting in a static color map, which has completely stable colors but poor discriminative power for large hierarchies.
The incremental colors, on the other hand, correspond to a stability ratio of $0$.
Here, the algorithm is unconstrained and allowed to freely distribute nodes in the color space to improve the discriminative power.
But this configuration yields larger differences in colors across interaction states, thus reducing color stability.

\input{interpolation.tex}

Based on these definitions, we can apply arbitrary stability ratios from $[0,1]$, to interpolate the desired tradeoff between the two quality criteria.
However, due to the cyclical axis, we can not interpolate hues naively.
As shown in \cref{fig:interpolation:shortestPath}, when varying the stability ratio between 0 and 1, each node follows the shortest path between its reference and incremental hue independently of the other nodes.
Thus, nodes may rotate in opposite directions, depending on which direction results in the shorter path.
In these cases, color maps for interpolated stability ratios do not accurately reflect the hierarchical relationships of the nodes.
To avoid these artifacts, we employ a hierarchical interpolation scheme, based on the offset between a node's hue and that of its parent (see \cref{fig:interpolation:referenceOffset}--\subref{fig:interpolation:incrementalOffset}).
We then interpolate between the incremental and reference offsets of each node and reconstruct the color values from the root of the hierarchy downwards, resulting in consistently interpolated color maps (see \cref{fig:interpolation:offsetPath}).

\section{\uppercase{Application Examples}}
\label{sec:examples}

In this section, we demonstrate the wide applicability of \dtc{ }with three application examples differing in their visual representation of the hierarchy and their filtering modality.
In the following sections, we use these examples as basis for our quality analysis and qualitative user study.
Throughout all examples we apply the static \tc{ }algorithm with a hue fraction of $0.75$~\citep{tennekesTreeColorsColor2014}.
We also implement a proportional hue split and a local interpolation, following the design rules specified in our previous work~\citep{mertzQualityApproachHierarchical2024}.

\paragraph*{Treemap.}

As first example, we replicate the zoomable treemap, shown in the original \cuttle{ }publication~\citep[Fig.~6]{waldinCuttlefishColorMapping2019}.
The dataset contains a hierarchy of 229 countries, divided into regions and sub-regions~\citep{ledozeWorldCountries2012}.
For each node, the population size is encoded by the rectangle area in the treemap (see \cref{fig:example:treemap}).
The treemap shows three hierarchy levels at once.
Initially, these are the root node that fills the entire viewport, the regions as rectangular areas with a thin border and the sub-regions as filled rectangles within the regions.
Users can zoom into the hierarchy by clicking into one of the regions.
This selects the clicked region as new root of the visualization, filling the entire viewport.
The sub-regions are now displayed as areas with a border, while the individual countries become visible as rectangles.
When clicking into one of the sub-regions, the view is zoomed in again.
But at this zoom-level only two hierarchy levels are visible, because countries have no descendant nodes.
Users can zoom out by clicking the bar at the top of the visualization.

\paragraph*{Node-Link Diagram.}

As counterpoint to the previous implicit hierarchy visualization with the semantic zoom, we chose a node-link diagram with a geometric zoom and pan for our second example.
\Cref{fig:example:nodelink} shows our recreation of an example graph from the \tc{ }publication~\citep[Fig.~2]{tennekesTreeColorsColor2014}.
We implement zooming with the mouse wheel and panning by clicking and dragging the mouse.
As the viewport changes, the incremental colors are recomputed, considering only sub-trees within the viewport.
This matches the interaction modality of the HIV example within the \cuttle{ }publication~\citep{waldinCuttlefishColorMapping2019}.

\paragraph*{Icicle Plot.}

The third example consists of an icicle plot, where sub-trees can be declared as invisible by clicking on their root nodes.
The \enquote{invisible} sub-trees are colored gray and incremental colors are recomputed for the remainder of the hierarchy.
\Cref{fig:example:icicle} applies this representation to the structure of the source code of the Amazon Prime Video Android application, which we extracted from the class names in the compiled android package.
To drill down into the relevant part of the code-base for a given analysis goal, users can successively apply filters, removing parts of the code-base they are not interested in.
For example, in the figure the user first excludes global namespaces other than the \emph{.com} namespace, thus removing many international libraries that are used within the application.
Then the user excludes the remaining non-Amazon namespaces, leaving only the internal code, developed by Amazon.
While successively applying filters and narrowing the analysis scope to the relevant parts of the hierarchy, the colors within the remaining hierarchy are updated to improve discriminative power.
We chose this interaction modality, again, as a contrast to the previous examples.
In this scenario, the layout of the icicle plot remains consistent and only the color map is adjusted.
In a real world application, this is useful if the the icicle plot is linked with other visualizations and used as a filtering widget to reduce the hierarchically structured data set.

\section{\uppercase{Color Map Quality}}

To provide empirical guidance for designers, we define quantitative metrics for the discriminative power and color stability in hierarchical color maps.
These metrics shall serve as an objective foundation for the discussion concerning the tradeoff between the two goals.
We then apply the metrics to investigate the attained quality from \dtc{ }and \cuttle.

\subsection{Quality Metrics}

We base our definitions of the metrics on those provided by \citet{bujackGoodBadUgly2018}.
However, their metrics are designed for continuous quantitative color maps with measurable differences between the represented values.
For hierarchical color maps, we must define metrics solely based on the color differences.
But we can reproduce the intent behind their definitions.

Our notation extends the one described in \cref{sec:algorithm}, such that $R$ is the reference-state, i.e. the entire hierarchy, and $S \subseteq R$ is a given interaction state, each described as the set of visible nodes.
We also define $n$ to refer to the number of nodes in the hierarchy, while $l$ is the number of leaf-nodes.
Alike \citet{bujackGoodBadUgly2018}, we denote $\Delta E$ as a hypothetical perfectly accurate color distance metric, and $x(v)$ as the color mapping function, that assigns a color value to the node $v$.
We also introduce a notation for sibling relationships with $a \curlywedge b$, indicating that the nodes $a$ and $b$ share the same parent.
For simplicity, we define that $a \curlywedge b$ implies $a \neq b$.

\paragraph*{Discriminative Power.}

Discriminative power measures the color-distances between nodes, thus indicating how distinguishable the generated colors are.
\citet{bujackGoodBadUgly2018} define a local and global discriminative power, where the first is constrained to neighboring samples along the color scale, while the latter measures distances between arbitrary samples.
We adapt this approach by considering pairwise color differences across the entire hierarchy (global) and among sibling nodes (local).
In contrast to \citet{bujackGoodBadUgly2018}, we do not have to deal with sampling artifacts, because the hierarchy is a discrete set of nodes, and we can simply take the average of all pairwise distances.
Hence, the \textbf{global discriminative power} is defined as:

\begin{equation}
     D_G (S) = \frac{\sum_{i \neq j} \Delta E(x_S(v_i), x_S(v_j))}{n^2}
\end{equation}

The \textbf{local discriminative power} is defined analogously, but only considers pair-wise differences between sibling nodes:

\begin{equation}
    D_L (S) = \frac{\sum_{v_i \curlywedge v_j} \Delta E(x_S(v_i), x_S(v_j))}{|\{(i, j): v_i \curlywedge v_j\}|}
\end{equation}

\paragraph*{Color Stability.}

The stability metric shall quantify the amount of color change that occurs during users' interaction with the visualization.
Hence, we define this metric in terms of color distances as well, resulting in a measure of color instability, where a lower value is preferable.
Here, we wish to point out that \dtc{ }and \cuttle{ }minimize different aspects of color instability.
\dtc{ }aligns colors with a static reference, while \cuttle{ }aligns with the colors of the previous interaction state.
As of today, it has not yet been investigated, which of the two aspects is more important for mental map preservation.
For this reason, we quantify both aspects in individual metrics.

The first metric is the \textbf{reference instability}, which captures the average amount of color change between a given interaction state and the reference color map (see \cref{eq:refStab}).
Ideally, one would compute the average reference instability over all possible interaction states of the hierarchy.
However, in the general case, where every possible combination of sub-trees can be visible, the number of possible states is in ${\bigO}(2^{l})$, thus sampling becomes necessary.

\begin{equation}
    I_R (S) = \frac{\sum_{v \in S} \Delta E(x_S (v), x_R (v))}{|S|}
    \label{eq:refStab}
\end{equation}

Furthermore, the \textbf{incremental instability} quantifies the average amount by which colors change when transitioning between two interaction states (see \cref{eq:incStab}).
Here, one would ideally compute the average over all possible pairs of interaction states, but due to the complexity concerns described above, it is more practical to sample a sufficiently large set of pairs.

\begin{equation}
    I_I (S_i, S_j) = \frac{\sum_{v \in (S_i \cap S_j)} \Delta E (x_{S_i} (v), x_{S_j} (v))}{|S_i \cap S_j|}
    \label{eq:incStab}
\end{equation}

\subsection{Experimental Setup}

After having defined four metrics to quantify different aspects of color map quality, we now apply these metrics to investigate the quality of color maps resulting from \dtc{ }and \cuttle.
Our first goal was to find out whether the interpolation approach via the stability ratio achieves its design goal by allowing the flexible adjustment of the tradeoff between discriminative power and color stability.
Second, we wanted to investigate whether the two algorithms' focus on minimizing the incremental or the reference instability are reflected in the results of the corresponding metric.

We applied \cuttle{ }and \dtc{ }with eleven equidistant stability ratios from $[0,1]$ to seven different data sets---the \emph{World Population} data set (see \cref{fig:example:treemap}), the node-link diagram example tree (see \cref{fig:example:nodelink}), a hierarchical structure of european countries, and four reduced code structures from different Android applications (\emph{Honda Connect}, \emph{Goldmessenger}, \emph{Prime Video}, \emph{Amazon Music}).
More details are available in the supplemental materials.
We approximate $\Delta E$ with the CIE $\Delta E_{2000}$ color distance metric~\citep{luoDevelopmentCIE20002001}, which, of current metrics, most closely resembles human perception.
To enable the comparison, we applied \cuttle{ }to the entire hierarchy, by iterating over the hierarchy levels and aligning the hues of each level with the previous.
We set the maximum angle between node-hues ($\beta_{max}$) to $5^{\circ}$ and the maximum angle between adjacent sub-trees ($\alpha_{max}$) to $30^{\circ}$, as specified in the treemap example within the \cuttle{ }publication~\citep{waldinCuttlefishColorMapping2019}.

\subsection{Results}

We sampled 10,000 random interaction states and state-pairs to compute our metrics.
In the following, we report each quality metric as a curve over the percentage of invisible nodes (for the sampled states) or nodes with changed visibility (for sampled state-pairs).
We discretize this percentage with equidistant bins and accumulate the average value of the metric for every bin.
To improve robustness against sampling artifacts, we discard bins with less than 10 samples.

\paragraph*{Discriminative Power.}

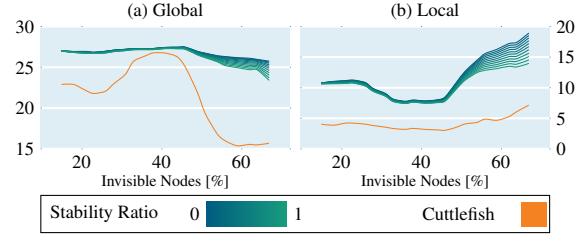
\begin{figure}
    \centering
    \scriptsize
    \begin{tikzpicture}
        \begin{groupplot}[
            group style={
                group size=2 by 1,
                horizontal sep=4pt,
                vertical sep=3em,
            },
            height=3.2cm,
            width=0.65\linewidth,
            tickwidth=0,
            axis line style={draw=none},
            axis background/.style={fill=fhgBlueLight},
            ymajorgrids,
            major grid style={thick,white},
            xlabel style={
                font=\tiny,
                inner sep=0pt
            }
        ]
            \nextgroupplot[
                xlabel={Invisible Nodes [\%]},
                yticklabel pos=left,
                ymin=15,
                ymax=30
            ]
                \addplot[draw=fhgBlue,smooth] coordinates{(15,27) (18,26.94) (21,26.95) (23,26.84) (26,26.97) (28,27.1) (31,27.21) (33,27.28) (36,27.3) (38,27.3) (41,27.44) (44,27.46) (46,27.47) (49,26.96) (51,26.73) (54,26.42) (56,26.32) (59,26.18) (62,26.11) (64,25.97) (67,25.75)};
                \addplot[draw=fhgBlue!90!fhg,smooth] coordinates{(15,27) (18,26.93) (21,26.93) (23,26.83) (26,26.95) (28,27.09) (31,27.20) (33,27.28) (36,27.29) (38,27.3) (41,27.43) (44,27.45) (46,27.45) (49,26.93) (51,26.7) (54,26.39) (56,26.25) (59,26.12) (62,26.01) (64,25.9) (67,25.64)};
                \addplot[draw=fhgBlue!80!fhg,smooth] coordinates{(15,27) (18,26.93) (21,26.91) (23,26.82) (26,26.94) (28,27.08) (31,27.19) (33,27.27) (36,27.28) (38,27.29) (41,27.42) (44,27.44) (46,27.44) (49,26.90) (51,26.65) (54,26.33) (56,26.16) (59,26.04) (62,25.94) (64,25.82) (67,25.5)};
                \addplot[draw=fhgBlue!70!fhg,smooth] coordinates{(15,27) (18,26.92) (21,26.88) (23,26.81) (26,26.92) (28,27.06) (31,27.17) (33,27.26) (36,27.27) (38,27.28) (41,27.41) (44,27.42) (46,27.42) (49,26.87) (51,26.61) (54,26.27) (56,26.07) (59,25.94) (62,25.82) (64,25.72) (67,25.31)};
                \addplot[draw=fhgBlue!60!fhg,smooth] coordinates{(15,27) (18,26.9) (21,26.86) (23,26.79) (26,26.9) (28,27.05) (31,27.16) (33,27.24) (36,27.26) (38,27.27) (41,27.4) (44,27.41) (46,27.39) (49,26.83) (51,26.55) (54,26.2) (56,25.98) (59,25.82) (62,25.68) (64,25.62) (67,25.11)};
                \addplot[draw=fhgBlue!50!fhg,smooth] coordinates{(15,27) (18,26.89) (21,26.83) (23,26.78) (26,26.87) (28,27.03) (31,27.14) (33,27.23) (36,27.25) (38,27.25) (41,27.38) (44,27.39) (46,27.37) (49,26.78) (51,26.5) (54,26.12) (56,25.86) (59,25.71) (62,25.55) (64,25.49) (67,24.89)};
                \addplot[draw=fhgBlue!40!fhg,smooth] coordinates{(15,26.99) (18,26.87) (21,26.81) (23,26.76) (26,26.85) (28,27.00) (31,27.12) (33,27.21) (36,27.23) (38,27.24) (41,27.37) (44,27.37) (46,27.34) (49,26.73) (51,26.44) (54,26.04) (56,25.74) (59,25.57) (62,25.40) (64,25.35) (67,24.62)};
                \addplot[draw=fhgBlue!30!fhg,smooth] coordinates{(15,26.99) (18,26.85) (21,26.78) (23,26.74) (26,26.82) (28,26.98) (31,27.1) (33,27.2) (36,27.22) (38,27.22) (41,27.35) (44,27.34) (46,27.31) (49,26.68) (51,26.37) (54,25.96) (56,25.62) (59,25.43) (62,25.24) (64,25.19) (67,24.33)};
                \addplot[draw=fhgBlue!20!fhg,smooth] coordinates{(15,26.99) (18,26.83) (21,26.75) (23,26.72) (26,26.79) (28,26.96) (31,27.07) (33,27.18) (36,27.2) (38,27.2) (41,27.33) (44,27.32) (46,27.27) (49,26.63) (51,26.31) (54,25.87) (56,25.49) (59,25.28) (62,25.06) (64,25.03) (67,24.03)};
                \addplot[draw=fhgBlue!10!fhg,smooth] coordinates{(15,26.99) (18,26.82) (21,26.72) (23,26.70) (26,26.76) (28,26.93) (31,27.05) (33,27.16) (36,27.18) (38,27.18) (41,27.31) (44,27.29) (46,27.24) (49,26.57) (51,26.23) (54,25.77) (56,25.34) (59,25.12) (62,24.9) (64,24.85) (67,23.71)};
                \addplot[draw=fhg,smooth] coordinates{(15,27) (18,26.83) (21,26.72) (23,26.71) (26,26.76) (28,26.92) (31,27.04) (33,27.15) (36,27.16) (38,27.17) (41,27.29) (44,27.27) (46,27.20) (49,26.52) (51,26.17) (54,25.67) (56,25.21) (59,24.96) (62,24.71) (64,24.69) (67,23.4)};
                \addplot[draw=fhgOrange,smooth] coordinates{(15,22.9) (18,22.9) (21,22.18) (23,21.79) (26,22.07) (28,22.95) (31,24.15) (33,25.72) (36,26.4) (38,26.77) (41,26.74) (44,26.27) (46,25.03) (49,21.95) (51,19.32) (54,16.83) (56,15.99) (59,15.38) (62,15.55) (64,15.49) (67,15.69)};
                \coordinate (global) at (rel axis cs:0.5,1);
                \coordinate (bottomLeft) at (rel axis cs:0,0);
            \nextgroupplot[
                xlabel={Invisible Nodes [\%]},
                yticklabel pos=right,
                ymin=0,
                ymax=20
            ]
                \addplot[draw=fhgBlue,smooth] coordinates{(15,10.77) (18,11.03) (21,11.15) (23,11.24) (26,10.84) (28,9.86) (31,9.07) (33,8.13) (36,7.78) (38,7.98) (41,7.84) (44,8.01) (46,8.47) (49,11) (51,12.66) (54,14.48) (56,15.56) (59,16.27) (62,17.21) (64,17.43) (67,18.93)};
                \addplot[draw=fhgBlue!90!fhg,smooth] coordinates{(15,10.77) (18,10.99) (21,11.10) (23,11.2) (26,10.79) (28,9.82) (31,9.03) (33,8.09) (36,7.74) (38,7.94) (41,7.80) (44,7.97) (46,8.41) (49,10.9) (51,12.53) (54,14.3) (56,15.31) (59,15.96) (62,16.89) (64,17.08) (67,18.54)};
                \addplot[draw=fhgBlue!80!fhg,smooth] coordinates{(15,10.76) (18,10.96) (21,11.06) (23,11.15) (26,10.73) (28,9.77) (31,8.98) (33,8.05) (36,7.70) (38,7.91) (41,7.76) (44,7.92) (46,8.34) (49,10.79) (51,12.39) (54,14.11) (56,15.06) (59,15.68) (62,16.54) (64,16.72) (67,18.13)};
                \addplot[draw=fhgBlue!70!fhg,smooth] coordinates{(15,10.75) (18,10.93) (21,11.01) (23,11.10) (26,10.67) (28,9.72) (31,8.94) (33,8.01) (36,7.67) (38,7.87) (41,7.73) (44,7.88) (46,8.28) (49,10.69) (51,12.25) (54,13.91) (56,14.8) (59,15.38) (62,16.19) (64,16.36) (67,17.68)};
                \addplot[draw=fhgBlue!60!fhg,smooth] coordinates{(15,10.73) (18,10.89) (21,10.97) (23,11.05) (26,10.62) (28,9.68) (31,8.89) (33,7.97) (36,7.63) (38,7.84) (41,7.69) (44,7.83) (46,8.21) (49,10.58) (51,12.10) (54,13.71) (56,14.53) (59,15.08) (62,15.82) (64,15.95) (67,17.21)};
                \addplot[draw=fhgBlue!50!fhg,smooth] coordinates{(15,10.72) (18,10.85) (21,10.92) (23,11.01) (26,10.56) (28,9.63) (31,8.85) (33,7.93) (36,7.59) (38,7.80) (41,7.65) (44,7.77) (46,8.14) (49,10.47) (51,11.96) (54,13.51) (56,14.26) (59,14.76) (62,15.46) (64,15.58) (67,16.73)};
                \addplot[draw=fhgBlue!40!fhg,smooth] coordinates{(15,10.69) (18,10.82) (21,10.87) (23,10.96) (26,10.5) (28,9.58) (31,8.80) (33,7.89) (36,7.55) (38,7.77) (41,7.61) (44,7.74) (46,8.08) (49,10.35) (51,11.81) (54,13.31) (56,13.99) (59,14.44) (62,15.08) (64,15.15) (67,16.22)};
                \addplot[draw=fhgBlue!30!fhg,smooth] coordinates{(15,10.68) (18,10.77) (21,10.82) (23,10.90) (26,10.44) (28,9.54) (31,8.76) (33,7.85) (36,7.52) (38,7.73) (41,7.57) (44,7.69) (46,8.01) (49,10.24) (51,11.67) (54,13.10) (56,13.71) (59,14.13) (62,14.68) (64,14.74) (67,15.64)};
                \addplot[draw=fhgBlue!20!fhg,smooth] coordinates{(15,10.66) (18,10.72) (21,10.77) (23,10.85) (26,10.38) (28,9.49) (31,8.71) (33,7.81) (36,7.48) (38,7.69) (41,7.53) (44,7.65) (46,7.94) (49,10.12) (51,11.52) (54,12.89) (56,13.43) (59,13.79) (62,14.3) (64,14.3) (67,15.10)};
                \addplot[draw=fhgBlue!10!fhg,smooth] coordinates{(15,10.63) (18,10.69) (21,10.73) (23,10.80) (26,10.31) (28,9.43) (31,8.66) (33,7.77) (36,7.44) (38,7.65) (41,7.49) (44,7.59) (46,7.86) (49,10.00) (51,11.36) (54,12.67) (56,13.13) (59,13.47) (62,13.91) (64,13.85) (67,14.54)};
                \addplot[draw=fhg,smooth] coordinates{(15,10.63) (18,10.65) (21,10.69) (23,10.76) (26,10.27) (28,9.4) (31,8.63) (33,7.74) (36,7.41) (38,7.63) (41,7.46) (44,7.56) (46,7.8) (49,9.89) (51,11.21) (54,12.46) (56,12.84) (59,13.12) (62,13.49) (64,13.39) (67,13.97)};
                \addplot[draw=fhgOrange,smooth] coordinates{(15,4.04) (18,3.86) (21,4.2) (23,4.21) (26,4.04) (28,3.81) (31,3.56) (33,3.33) (36,3.19) (38,3.35) (41,3.18) (44,3.11) (46,3.03) (49,3.57) (51,4.07) (54,4.33) (56,4.86) (59,4.67) (62,5.27) (64,6.1) (67,7.16)};
                \coordinate (local) at (rel axis cs:0.5,1);
                \coordinate (bottomRight) at (rel axis cs:1,0);
        \end{groupplot}
        \node[anchor=south] at (global) {\phantomsubcaption \label{fig:discriminativePowerRandom:global}(a) Global};
        \node[anchor=south] at (local) {\phantomsubcaption \label{fig:discriminativePowerRandom:local}(b) Local};
        \draw ($(bottomLeft)+(0,-2.5em)$) rectangle ($(bottomRight)+(0,-4.5em)$);
        \node[anchor=west] (legendFirst) at ($(bottomLeft)+(1pt,-3.5em)$) {Stability Ratio};
        \node[anchor=west] (legendSecond) at ($(legendFirst.east)+(1em,0)$) {0};
        \node[anchor=west,left color=fhgBlue,right color=fhg] (legendThird) at (legendSecond.east) {\phantom{XXXXX}};
        \node[anchor=west] (legendFourth) at (legendThird.east) {1};
        \node[anchor=east,fill=fhgOrange] (legendSixth) at ($(bottomRight)+(-1pt,-3.5em)$) {\phantom{X}};
        \node[anchor=east] (legendFifth) at ($(legendSixth.west)+(-1em,0)$) {Cuttlefish};
    \end{tikzpicture}
    \caption{Discriminative power averaged over all data sets.}
    \label{fig:discriminativePowerRandom}
\end{figure}

\Cref{fig:discriminativePowerRandom} shows the results for the discriminative power over all data sets.
Notably, the global discriminative power decreases in all conditions as a larger fraction of nodes becomes invisible.
This is counter-intuitive, because the dynamic recomputation of colors is explicitly designed to increase discriminability.
This trend is caused by the metric considering both distances within sub-trees (such as a node and its ancestors) and between sub-trees.
The former distances are smaller than the latter, but as more sub-trees become invisible, the number of between-sub-tree pairs decreases more rapidly than the number of within-sub-tree pairs.
Thus, the relative weight of the smaller within-sub-tree distances increases, and the global discriminative power decreases.
The local discriminative power does not suffer from this phenomenon, because it only considers differences between sibling nodes.
Thus, the local metric shows and increasing trend as expected.

The curves of the individual stability ratios indicate the expected relationship between this parameter and the resulting discriminative power.
In both metrics, the curves are tightly bundled for low fractions of invisible nodes and diverge as the fraction increases.
The equidistant stability ratios also result in equidistant curves.
Furthermore, higher stability ratios result in a lower discriminative power.

Finally, \cuttle{ }achieves consistently lower discriminative power for both metrics than any configuration of \dtc.
But an investigation into individual data sets provides more nuanced insights.
We report the global discriminative power for the \emph{World Population} and \emph{Goldmessenger} data sets in \cref{fig:discriminativePowerComparison}.

\input{QDiscriminativePowerComparison.tex}

\begin{figure}
    \centering
    \begin{subfigure}{0.7\linewidth}
        \centering
        \includegraphics[width=0.8\linewidth]{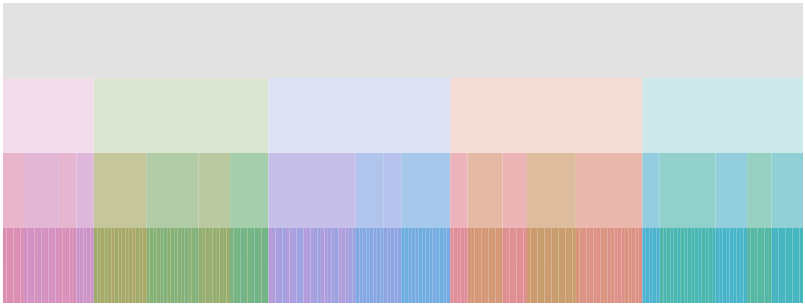}
    \end{subfigure}
    \begin{subfigure}{0.28\linewidth}
        \centering
        \includegraphics[width=0.8\linewidth]{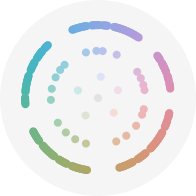}
    \end{subfigure}
    \begin{subfigure}{0.7\linewidth}
        \centering
        \includegraphics[width=0.8\linewidth]{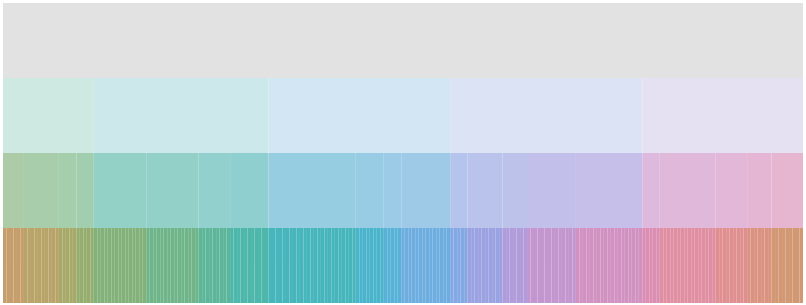}
    \end{subfigure}
    \begin{subfigure}{0.28\linewidth}
        \centering
        \includegraphics[width=0.8\linewidth]{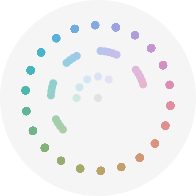}
    \end{subfigure}
    \caption{The World Population dataset, colored with Tree Colors (top row) and Cuttlefish (bottom row). The right side shows the distribution of nodes on the color wheel.}
    \label{fig:worldPopulation_comparison}
\end{figure}

\begin{figure}
    \centering
    \begin{subfigure}{0.7\linewidth}
        \centering
        \includegraphics[width=0.8\linewidth]{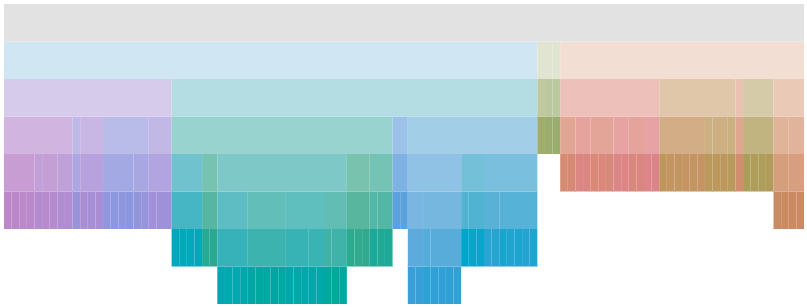}
    \end{subfigure}
    \begin{subfigure}{0.28\linewidth}
        \centering
        \includegraphics[width=0.8\linewidth]{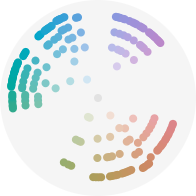}
    \end{subfigure}
    \begin{subfigure}{0.7\linewidth}
        \centering
        \includegraphics[width=0.8\linewidth]{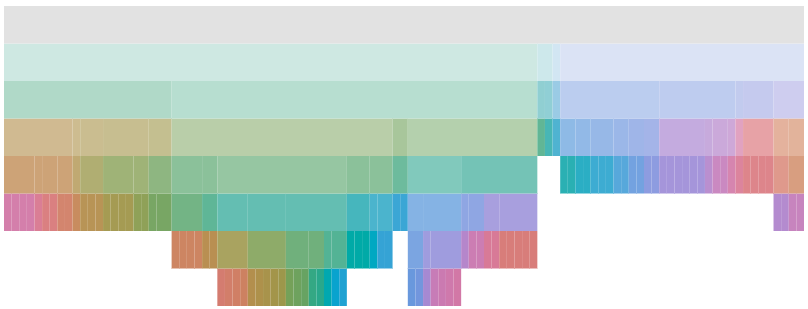}
    \end{subfigure}
    \begin{subfigure}{0.28\linewidth}
        \centering
        \includegraphics[width=0.8\linewidth]{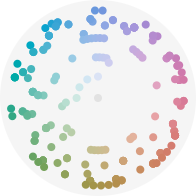}
    \end{subfigure}
    \caption{The Goldmessenger dataset, colored with Tree Colors (top row) and Cuttlefish (bottom row). The right side shows the distribution of nodes on the color wheel.}
    \label{fig:goldmessenger_comparison}
\end{figure}

In the \emph{World Population} dataset (\cref{fig:discriminativePowerComparison:worldPop}), \cuttle{ }achieves a lower global discriminative power than \dtc, while it achieves a higher score for the \emph{Goldmessenger} dataset.
To investigate the cause of this discrepancy, we display results of both algorithms for the \emph{World Population} dataset in \cref{fig:worldPopulation_comparison} and for the \emph{Goldmessenger} dataset in \cref{fig:goldmessenger_comparison}.
We can see that both algorithms utilize the full range of hues for the leaf nodes of the \emph{World Population} dataset.
However, for the inner nodes, \cuttle{ }utilizes a smaller range than \tc, due to the constraint of the $\alpha_{max}$ and $\beta_{max}$ parameters.
In the same figure, we can also observe that the recursive subdivision approach of \tc{ }propagates gaps between siblings downwards through the hierarchy levels, while \cuttle{ }computes positions and sizes of gaps in the hue distribution independently for the hierarchy levels.

For a less balanced dataset, such as the \emph{Goldmessenger} dataset (\cref{fig:goldmessenger_comparison}), this effect becomes more pronounced.
We can see that \tc{ }creates large gaps between the sub-trees that branch on the first level of the hierarchy.
This results in a comparatively large range of hues that is not utilized.
\cuttle, on the other hand, manages to distribute the nodes more evenly over the entire range of hues, thus, achieving an overall greater discriminative power.
But the independent gap-size computation of \cuttle{ }causes a different issue.
Looking at the icicle plot, we can see the same hues occurring in entirely unrelated parts of the hierarchy, but on different hierarchy levels.
These inconsistencies were not relevant for \cuttle's application to multi-scale visualizations, because only a single hierarchy level is visible at a time.
But applying the algorithm to general hierarchical visualizations yields results that arguably do not reflect the hierarchical structure.
The discriminative power does not capture these inconsistencies, because it only measures the magnitude of color differences.
So, while \cuttle{ }achieves a higher global discriminative power for such imbalanced datasets, its uniformity, i.e. its adherence to the hierarchical structure~\citep{mertzQualityApproachHierarchical2024}, is worse.
But at this time, we do not possess a quantifiable metric for uniformity in hierarchical color maps.

\paragraph*{Color Instability.}

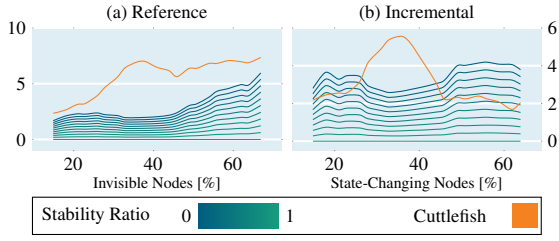
\begin{figure}
    \centering
    \scriptsize
    \begin{tikzpicture}
        \begin{groupplot}[
            group style={
                group size=2 by 1,
                horizontal sep=4pt,
                vertical sep=3em,
            },
            height=3.2cm,
            width=0.65\linewidth,
            tickwidth=0,
            axis line style={draw=none},
            axis background/.style={fill=fhgBlueLight},
            ymajorgrids,
            major grid style={thick,white},
            xlabel style={
                font=\tiny,
                inner sep=0pt
            }
        ]
            \nextgroupplot[
                xlabel={Invisible Nodes [\%]},
                yticklabel pos=left,
                ymin=-1,
                ymax=10
            ]
                \addplot[draw=fhgBlue,smooth] coordinates{(15,1.68) (18,2.10) (21,2.30) (23,2.27) (26,2.38) (28,2.22) (31,2.15) (33,1.96) (36,1.96) (38,1.98) (41,2.04) (44,2.17) (46,2.40) (49,3.03) (51,3.20) (54,3.79) (56,4.32) (59,4.58) (62,4.92) (64,4.89) (67,5.97)};
                \addplot[draw=fhgBlue!90!fhg,smooth] coordinates{(15,1.51) (18,1.89) (21,2.06) (23,2.05) (26,2.15) (28,2) (31,1.93) (33,1.76) (36,1.77) (38,1.78) (41,1.84) (44,1.95) (46,2.17) (49,2.74) (51,2.89) (54,3.41) (56,3.9) (59,4.14) (62,4.44) (64,4.41) (67,5.41)};
                \addplot[draw=fhgBlue!80!fhg,smooth] coordinates{(15,1.35) (18,1.69) (21,1.84) (23,1.83) (26,1.91) (28,1.78) (31,1.72) (33,1.57) (36,1.57) (38,1.59) (41,1.63) (44,1.74) (46,1.93) (49,2.43) (51,2.57) (54,3.04) (56,3.47) (59,3.68) (62,3.95) (64,3.93) (67,4.83)};
                \addplot[draw=fhgBlue!70!fhg,smooth] coordinates{(15,1.18) (18,1.48) (21,1.62) (23,1.6) (26,1.67) (28,1.56) (31,1.51) (33,1.38) (36,1.38) (38,1.39) (41,1.43) (44,1.52) (46,1.69) (49,2.13) (51,2.25) (54,2.66) (56,3.04) (59,3.22) (62,3.46) (64,3.44) (67,4.25)};
                \addplot[draw=fhgBlue!60!fhg,smooth] coordinates{(15,1.02) (18,1.27) (21,1.39) (23,1.37) (26,1.44) (28,1.34) (31,1.3) (33,1.18) (36,1.18) (38,1.2) (41,1.23) (44,1.31) (46,1.45) (49,1.83) (51,1.93) (54,2.29) (56,2.61) (59,2.77) (62,2.97) (64,2.96) (67,3.66)};
                \addplot[draw=fhgBlue!50!fhg,smooth] coordinates{(15,0.85) (18,1.06) (21,1.16) (23,1.15) (26,1.2) (28,1.12) (31,1.08) (33,0.99) (36,0.99) (38,1.00) (41,1.03) (44,1.09) (46,1.21) (49,1.53) (51,1.61) (54,1.91) (56,2.18) (59,2.31) (62,2.48) (64,2.47) (67,3.05)};
                \addplot[draw=fhgBlue!40!fhg,smooth] coordinates{(15,0.68) (18,0.85) (21,0.93) (23,0.92) (26,0.96) (28,0.9) (31,0.87) (33,0.79) (36,0.8) (38,0.80) (41,0.83) (44,0.88) (46,0.97) (49,1.23) (51,1.29) (54,1.53) (56,1.75) (59,1.85) (62,1.98) (64,1.99) (67,2.44)};
                \addplot[draw=fhgBlue!30!fhg,smooth] coordinates{(15,0.52) (18,0.64) (21,0.70) (23,0.69) (26,0.72) (28,0.68) (31,0.66) (33,0.60) (36,0.60) (38,0.61) (41,0.63) (44,0.67) (46,0.74) (49,0.92) (51,0.98) (54,1.15) (56,1.32) (59,1.4) (62,1.5) (64,1.49) (67,1.84)};
                \addplot[draw=fhgBlue!20!fhg,smooth] coordinates{(15,0.36) (18,0.43) (21,0.47) (23,0.47) (26,0.49) (28,0.46) (31,0.44) (33,0.41) (36,0.41) (38,0.41) (41,0.42) (44,0.45) (46,0.5) (49,0.62) (51,0.66) (54,0.78) (56,0.88) (59,0.93) (62,1.00) (64,1.00) (67,1.23)};
                \addplot[draw=fhgBlue!10!fhg,smooth] coordinates{(15,0.19) (18,0.22) (21,0.24) (23,0.24) (26,0.25) (28,0.23) (31,0.23) (33,0.21) (36,0.21) (38,0.22) (41,0.22) (44,0.23) (46,0.26) (49,0.32) (51,0.34) (54,0.4) (56,0.45) (59,0.47) (62,0.51) (64,0.51) (67,0.62)};
                \addplot[draw=fhg,smooth] coordinates{(15,0) (18,0) (21,0) (23,0) (26,0) (28,0) (31,0) (33,0) (36,0) (38,0) (41,0) (44,0) (46,0) (49,0) (51,0) (54,0) (56,0) (59,0) (62,0) (64,0) (67,0)};
                \addplot[draw=fhgOrange,smooth] coordinates{(15,2.35) (18,2.63) (21,3.14) (23,3.22) (26,3.92) (28,4.69) (31,5.56) (33,6.43) (36,6.95) (38,7) (41,6.47) (44,6.17) (46,5.63) (49,6.37) (51,6.39) (54,6.82) (56,6.82) (59,7.08) (62,6.98) (64,6.89) (67,7.36)};
                \coordinate (global) at (rel axis cs:0.5,1);
                \coordinate (bottomLeft) at (rel axis cs:0,0);
            \nextgroupplot[
                xlabel={State-Changing Nodes [\%]},
                yticklabel pos=right,
                ymin=-0.5,
                ymax=6
            ]
                \addplot[draw=fhgBlue,smooth] coordinates{(15,2.8) (18,3.65) (21,3.29) (23,3.47) (26,3.44) (28,2.94) (31,2.70) (33,2.58) (36,2.65) (38,2.74) (41,2.88) (44,3.05) (46,3.22) (49,3.99) (51,4.02) (54,4.14) (56,4.19) (59,4.08) (62,4.06) (64,3.79)};
                \addplot[draw=fhgBlue!90!fhg,smooth] coordinates{(15,2.51) (18,3.29) (21,2.97) (23,3.13) (26,3.09) (28,2.65) (31,2.43) (33,2.32) (36,2.39) (38,2.47) (41,2.6) (44,2.75) (46,2.90) (49,3.6) (51,3.63) (54,3.73) (56,3.78) (59,3.68) (62,3.66) (64,3.42)};
                \addplot[draw=fhgBlue!80!fhg,smooth] coordinates{(15,2.24) (18,2.94) (21,2.65) (23,2.79) (26,2.75) (28,2.36) (31,2.17) (33,2.07) (36,2.13) (38,2.2) (41,2.31) (44,2.45) (46,2.58) (49,3.20) (51,3.22) (54,3.32) (56,3.36) (59,3.27) (62,3.24) (64,3.02)};
                \addplot[draw=fhgBlue!70!fhg,smooth] coordinates{(15,1.96) (18,2.57) (21,2.32) (23,2.44) (26,2.42) (28,2.07) (31,1.9) (33,1.81) (36,1.87) (38,1.92) (41,2.03) (44,2.14) (46,2.26) (49,2.81) (51,2.82) (54,2.91) (56,2.95) (59,2.86) (62,2.84) (64,2.66)};
                \addplot[draw=fhgBlue!60!fhg,smooth] coordinates{(15,1.67) (18,2.21) (21,1.99) (23,2.09) (26,2.07) (28,1.78) (31,1.63) (33,1.56) (36,1.60) (38,1.65) (41,1.74) (44,1.84) (46,1.94) (49,2.41) (51,2.43) (54,2.50) (56,2.53) (59,2.45) (62,2.43) (64,2.3)};
                \addplot[draw=fhgBlue!50!fhg,smooth] coordinates{(15,1.41) (18,1.85) (21,1.66) (23,1.75) (26,1.73) (28,1.48) (31,1.36) (33,1.30) (36,1.34) (38,1.38) (41,1.45) (44,1.54) (46,1.62) (49,2.01) (51,2.02) (54,2.08) (56,2.12) (59,2.05) (62,2.03) (64,1.92)};
                \addplot[draw=fhgBlue!40!fhg,smooth] coordinates{(15,1.14) (18,1.48) (21,1.33) (23,1.40) (26,1.39) (28,1.19) (31,1.09) (33,1.04) (36,1.07) (38,1.11) (41,1.17) (44,1.23) (46,1.30) (49,1.61) (51,1.62) (54,1.67) (56,1.69) (59,1.63) (62,1.62) (64,1.52)};
                \addplot[draw=fhgBlue!30!fhg,smooth] coordinates{(15,0.85) (18,1.12) (21,1.01) (23,1.06) (26,1.04) (28,0.90) (31,0.83) (33,0.79) (36,0.81) (38,0.84) (41,0.88) (44,0.93) (46,0.98) (49,1.21) (51,1.22) (54,1.25) (56,1.27) (59,1.23) (62,1.22) (64,1.15)};
                \addplot[draw=fhgBlue!20!fhg,smooth] coordinates{(15,0.57) (18,0.74) (21,0.68) (23,0.71) (26,0.70) (28,0.61) (31,0.56) (33,0.53) (36,0.55) (38,0.57) (41,0.59) (44,0.63) (46,0.66) (49,0.82) (51,0.82) (54,0.84) (56,0.85) (59,0.84) (62,0.81) (64,0.76)};
                \addplot[draw=fhgBlue!10!fhg,smooth] coordinates{(15,0.28) (18,0.38) (21,0.35) (23,0.36) (26,0.36) (28,0.31) (31,0.29) (33,0.28) (36,0.28) (38,0.29) (41,0.31) (44,0.33) (46,0.34) (49,0.42) (51,0.42) (54,0.43) (56,0.44) (59,0.43) (62,0.41) (64,0.39)};
                \addplot[draw=fhg,smooth] coordinates{(15,0) (18,0) (21,0) (23,0) (26,0) (28,0) (31,0) (33,0) (36,0) (38,0) (41,0) (44,0) (46,0) (49,0) (51,0) (54,0) (56,0) (59,0) (62,0) (64,0)};
                \addplot[draw=fhgOrange,smooth] coordinates{(15,2.18) (18,2.45) (21,2.55) (23,2.56) (26,3.14) (28,4.16) (31,4.87) (33,5.38) (36,5.56) (38,5.20) (41,4.19) (44,3.13) (46,2.23) (49,2.33) (51,2.24) (54,2.39) (56,2.25) (59,2.07) (62,1.68) (64,2.05)};
                \coordinate (local) at (rel axis cs:0.5,1);
                \coordinate (bottomRight) at (rel axis cs:1,0);
        \end{groupplot}
        \node[anchor=south] at (global) {\phantomsubcaption \label{fig:stabilityRandom:reference}(a) Reference};
        \node[anchor=south] at (local) {\phantomsubcaption \label{fig:stabilityRandom:incremental}(b) Incremental};
        \draw ($(bottomLeft)+(0,-2.5em)$) rectangle ($(bottomRight)+(0,-4.5em)$);
        \node[anchor=west] (legendFirst) at ($(bottomLeft)+(1pt,-3.5em)$) {Stability Ratio};
        \node[anchor=west] (legendSecond) at ($(legendFirst.east)+(1em,0)$) {0};
        \node[anchor=west,left color=fhgBlue,right color=fhg] (legendThird) at (legendSecond.east) {\phantom{XXXXX}};
        \node[anchor=west] (legendFourth) at (legendThird.east) {1};
        \node[anchor=east,fill=fhgOrange] (legendSixth) at ($(bottomRight)+(-1pt,-3.5em)$) {\phantom{X}};
        \node[anchor=east] (legendFifth) at ($(legendSixth.west)+(-1em,0)$) {Cuttlefish};
    \end{tikzpicture}
    \caption{Color instability averaged over all data sets.}
    \label{fig:stabilityRandom}
\end{figure}

\Cref{fig:stabilityRandom} shows the color instability over all data sets.
For both metrics, we observe similar relationships as for the discriminative power.
As the fraction of invisible nodes grows, the incremental colors diverge from the reference color map, increasing the reference instability.
Decreasing the stability ratio, also increases instability, while a stability ratio of $1$ results in a static color map with a constant instability of $0$.
Overall, \cuttle{ }results in a greater reference instability than \dtc{ }, while the incremental instability is not as easy to judge.
Looking at individual data sets, we can, again, observe that \cuttle{ }achieves comparable incremental instability for balanced hierarchies (\cref{fig:stabilityComparison:worldPop}), while it yields larger instability for imbalanced hierarchies (\cref{fig:stabilityComparison:goldmessenger}).

\input{QInstabilityComparison.tex}

\paragraph*{Treemap Sampling.}

However, random sampling of arbitrary interaction states does not represent the application scenario that \cuttle{ }was designed for.
\cuttle{ }was designed with multi-scale visualizations in mind.
To facilitate a fair comparison, we investigate the treemap example from \cref{sec:examples}, and change our sampling strategy to incorporate only those states and state-transitions that are reachable with the semantic zooming interaction.
Because the number of possible states and state-transitions in this scenario is in $\bigO(n)$, we include all of them in our computation.
Furthermore, we only compute color distances between nodes on the visible hierarchy levels in the given interaction state.
The results from this adjusted computation (see \cref{fig:treemapSampling}) contain much larger oscillations in the curves, because of the smaller number of samples.
We can see that \cuttle{ }overall achieves a lower discriminative power than \dtc.
In terms of reference instability, the results for \cuttle, fall within the range of those covered by the various stability ratios.
Notably, \cuttle{ }achieves a very low incremental instability, which indicates that \cuttle{ }does perform well in its intended scenario.

\input{QTreemap.tex}

\section{\uppercase{User Study}}
\label{sec:userstudy}

To investigate the behavior of the two algorithms and the quality metrics in practice, we performed an exploratory user study.
Our main goals were to provide initial insights towards a verification of the measured relationships between stability ratio and the quality metrics as well as the identification of interesting research challenges for future studies.
To investigate the results of our quality analysis from the user perspective, we formed the following hypotheses:

\begin{description}
    \item[H1] A high stability ratio reduces the amount of user confusion due to color changes.
    \item[H2] A low stability ratio improves users' capability to identify colors.
    \item[H3] Depending on the analysis scenario, the optimal stability ratio differs.
    \item[H4] \dtc{ }with the optimal stability ratio for a given scenario achieves a better tradeoff between discriminability and stability than \cuttle.
\end{description}

By verifying the first two hypotheses, we can verify that our algorithm works as intended from the user perspective, and that our quality metrics measure what they are supposed to.
For the third hypothesis, we introduce our three application examples from \cref{sec:examples} as experimental variable.
We hypothesize, that with different visual encodings, interaction methods, and analysis tasks, the requirements for the tradeoff between discriminability and stability differ.
To that end, we adjusted the application examples to generate three different combinations of visualization type, interaction method, and task, which we call the three analysis scenarios.
Finally, we hypothesize that the flexibility of the stability ratio adjustment introduced by \dtc, allows to fine-tune the tradeoff to the analysis scenario, yielding superior performance to \cuttle{ }in all three scenarios when tuned appropriately.

\subsection{Experimental Variables}

We measure both discriminability and user confusion due to color changes as subjective ratings on a five-point Likert scale (\emph{1 - Very Little}, \emph{2 - Little}, \emph{3 - Acceptable}, \emph{4 - Much}, \emph{5 - Very Much}).
As independent variables, we investigate the three analysis scenarios and four different color maps---\cuttle{ }and \dtc{ }with three different stability ratios.
We sample three equidistant stability ratios throughout the spectrum, yielding one configuration with low stability ratio (0), one with medium stability ratio (0.4), and one with high stability ratio (0.8).
We avoided utilizing the entire spectrum, because a stability ratio of 1 results in a static color map, which would not provide useful insight into the subjective perception of color stability.

As stated above, we adjusted the individual application examples to implement different tasks that require users to identify hierarchy nodes based on their color, then change the interaction state and re-identify the same nodes again.
With these designs users were forced into situations where they would have to rely on the color map to identify nodes based on the color, only for that color to change after their next interaction.
Through this design users were able to subjectively assess how well they were supported in the identification of the nodes they were looking for, and how much the color changes were confusing them.

We adapted the treemap scenario by removing the labels of the lowest visible hierarchy level.
We only displayed the label of the root-level in the top bar and of the intermediate level within the corresponding area.
The lowest hierarchy level was exclusively identifiable via a color legend below the treemap.
In this scenario, users were tasked with comparing the population sizes of two countries that were in the same region, but not in the same sub-region.
Thus, users were required to zoom into the given region, to perform the comparison.
However, when zoomed into a region, the colors of the individual countries are not distinct enough to allow identification, so users had to zoom in to the individual sub-regions to identify the country.
After having identified the countries in their respective sub-regions, users had to zoom back out to view the entire region and perform the comparison.

Inspired by the application examples from the \tc{ }publication~\citep{tennekesTreeColorsColor2014}, we adapted the node-link scenario to display financial statistics of hierarchically organized entities.
We utilized a publicly available data set containing the net turnover development index of the retail industry in 35 European countries~\citep{eurostatRetailTradeTurnover2025}.
We organized the countries hierarchically, by first splitting the data set into members of the European Union and non-members.
We then further divided the EU members into countries of the Euro Area and countries where a different currency is used.

During the study, the node-link diagram displayed this hierarchical structure of countries, but only showed the individual nodes' labels, when users were fully zoomed in.
We also displayed the actual financial data in a line chart below the node-link diagram.
The line chart displayed the lines in the same color as their corresponding nodes in the node-link diagram, and hid those countries that were outside of the node-link diagram's viewport.
Participants were tasked with comparing the lines of two given countries.
In this scenario, participants were forced to zoom in very closely to reveal the labels of the individual countries.
After having found both countries of interest, users had to adjust the viewport to include both countries and as few other countries as possible.
During this viewport adjustment users were subjected to color changes.
Finally, users needed to identify the lines in the line chart that correspond to the given countries, to perform the comparison.

For the icicle plot scenario, we extracted sequences of instructions from four different android applications, using FlowDroid~\citep{arztFlowDroidPreciseContext2014}, and showed participants the hierarchical structure of the classes that contain these instructions.
Below the icicle plot that represented this hierarchical structure, we displayed a sequence diagram of the instruction sequences.
Each instruction in the sequence diagram was colored corresponding to its class in the hierarchy.
As filters were applied to the icicle plot, the sequence diagram was updated correspondingly.
For each of the four data sets, participants were then tasked to identify the class whose instructions occur most frequently in the sequence diagram.
To that end, participants had to successively apply filters to the hierarchy, to make it easier to distinguish the remaining instructions in the sequence diagram.

\subsection{Study Procedure}

We performed the study in a controlled environment with neutral lighting.
We started with a briefing, introducing the concept of dynamic hierarchical color maps as well as the opposing design goals of discriminative power and color stability.
Afterwards, we described the study procedure and presented participants with the study terms, asking for their consent.
Participants then had to fill out a short demographic questionnaire, after which the main part of the study started.
For this part, we pseudo-randomly assigned each participant two of the three scenarios.
Within each scenario, we started with an explanation of the visualization, the dataset and the interaction modality, followed by a training task, to learn the visualization and interaction.
Then, participants had to solve four trial tasks, as described above, each with a different color map.
To reduce biases due to varying trial difficulties, we kept the prompts in the same order for all participants, but randomized the order of the applied color map.
Randomization was achieved using a latin-square with four rows and columns.
In line with the exploratory nature of the study, we focused on gathering qualitative feedback and subjective ratings.
To that end, participants were asked to subjectively rate the perceived discriminative power and the level of confusion introduced by color instability after each task.
We observed participants during their work on the tasks to note their comments and the difficulties they encountered.
We were also available to answer questions or provide hints if they were having difficulties.
After completing both scenarios, we asked participants, whether they wanted to complete the third scenario as well, but we made it clear, that this was optional.
Finally, we gave participants the opportunity to provide further comments on the study.

\subsection{Results}

We recruited 18 participants from employees and students of our research institution and from the authors' personal network.
To avoid the social-desirability bias induced by this recruitment strategy, we only briefed participants on our goal to compare different color maps.
We disclosed our authorship of \dtc{ }only after the study was complete.
Each study took 30 to 45 minutes to complete the first two scenarios, and 45 to 75 minutes to complete all three scenarios, which all but one participant agreed to.
The demographic distribution is shown in \cref{fig:demographics}.
No participants reported any color vision deficiency.

\begin{figure}
    \centering
    \scriptsize
    \begin{tikzpicture}
        \begin{groupplot}[
            group style={
                group size=2 by 2,
                horizontal sep=4pt,
                vertical sep=3em,
                yticklabels at=edge left
            },
            height=2.5cm,
            width=0.65\linewidth,
            ybar,
            xtick={data},
            tickwidth=0,
            ymin=0,
            ymax=15,
            axis line style={draw=none},
            axis background/.style={fill=fhgBlueLight},
            ymajorgrids,
            major grid style={thick,white},
        ]
            \nextgroupplot[
                enlarge x limits=0.2,
                xticklabels={20--29, 30--39, 40--49, 50--59}
            ]
                \addplot[fill=fhgBlue,draw=none] coordinates{(0,7) (1,7) (2,3) (3,1)};
                \coordinate (age) at (rel axis cs:0.5,1);
            \nextgroupplot[
                enlarge x limits=0.5,
                xticklabels={Female, Male}
            ]
                \addplot[fill=fhgBlue,draw=none] coordinates{(0,7) (1,11)};
                \coordinate (gender) at (rel axis cs:0.5,1);
            \nextgroupplot[
                enlarge x limits=0.2,
                xticklabels={A-Levels, Apprenticeship, University, Doctorate},
                x tick label style={
                    rotate=-90,
                    anchor=west
                }
            ]
                \addplot[fill=fhgBlue,draw=none] coordinates{(0,2) (1,2) (2,12) (3,2)};
                \coordinate (degree) at (rel axis cs:0.5,1);
            \nextgroupplot[
                xticklabels={1 - Very Little, 2 - Little, 3 - Acceptable, 4 - Much, 5 - Very Much},
                x tick label style={
                    rotate=-90,
                    anchor=west
                }
            ]
                \addplot[fill=fhgBlue,draw=none] coordinates{(1,2) (2,1) (3,7) (4,4) (5,4)};
                \coordinate (comp_vis) at (rel axis cs:0.5,1);
        \end{groupplot}
        \node[anchor=south] at (age) {Age};
        \node[anchor=south] at (gender) {Gender};
        \node[anchor=south] at (degree) {Education};
        \node[anchor=south] at (comp_vis) {Visualization Literacy};
    \end{tikzpicture}
    \caption{Demographic distribution of study participants.}
    \label{fig:demographics}
\end{figure}
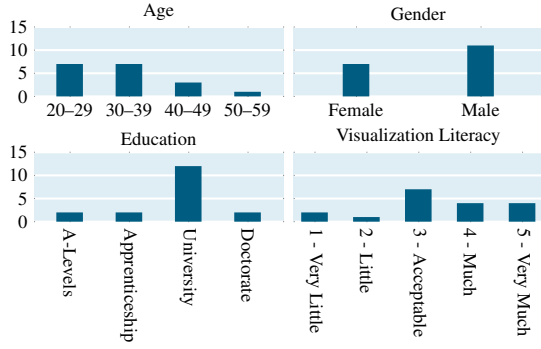

\begin{figure}
    \scriptsize
    \centering
    \begin{subfigure}{\linewidth}
        \centering
        \begin{tikzpicture}
            \begin{groupplot}[
                group style={
                    group size=3 by 3,
                    horizontal sep=4pt,
                    vertical sep=2pt,
                    yticklabels at=edge left,
                    xticklabels at=edge bottom,
                    xlabels at=edge top
                },
                width=0.49\linewidth,
                ytick={data},
                yticklabels={0, 0.4, 0.8, CF},
                yticklabel style={
                    align=right
                },
                tickwidth=0,
                xmin=0,
                xmax=18,
                xlabel shift={-1em},
                axis line style={draw=none},
                y dir=reverse,
                enlarge y limits=0.2,
                xbar stacked
            ]
                \nextgroupplot[bar width=1.5em, ylabel={\phantomsubcaption \label{fig:UResults:DiscriminativePower} (a) Discr. Power}, xlabel={Treemap}]
                    \addplot[fill=fhgBlue,draw=none] coordinates{(1,0) (4,1) (5,2) (1,3)};
                    \addplot[fill=fhgBlue!66,draw=none] coordinates{(3,0) (3,1) (9,2) (4,3)};
                    \addplot[fill=black!20,draw=none] coordinates{(3,0) (5,1) (4,2) (4,3)};
                    \addplot[fill=fhgOrange!66,draw=none] coordinates{(5,0) (5,1) (0,2) (7,3)};
                    \addplot[fill=fhgOrange,draw=none] coordinates{(6,0) (1,1) (0,2) (2,3)};
                \nextgroupplot[bar width=1.5em, xlabel={Node-Link Diagram}]
                    \addplot[fill=fhgBlue,draw=none] coordinates{(1,0) (1,1) (2,2) (8,3)};
                    \addplot[fill=fhgBlue!66,draw=none] coordinates{(2,0) (4,1) (6,2) (5,3)};
                    \addplot[fill=black!20,draw=none] coordinates{(3,0) (3,1) (5,2) (3,3)};
                    \addplot[fill=fhgOrange!66,draw=none] coordinates{(6,0) (9,1) (4,2) (1,3)};
                    \addplot[fill=fhgOrange,draw=none] coordinates{(6,0) (1,1) (1,2) (1,3)};
                \nextgroupplot[bar width=1.5em, xlabel={Icicle Plot}]
                    \addplot[fill=fhgBlue,draw=none] coordinates{(0,0) (0,1) (1,2) (3,3)};
                    \addplot[fill=fhgBlue!66,draw=none] coordinates{(2,0) (1,1) (4,2) (4,3)};
                    \addplot[fill=black!20,draw=none] coordinates{(1,0) (3,1) (4,2) (2,3)};
                    \addplot[fill=fhgOrange!66,draw=none] coordinates{(11,0) (8,1) (2,2) (3,3)};
                    \addplot[fill=fhgOrange,draw=none] coordinates{(3,0) (5,1) (6,2) (5,3)};
                \nextgroupplot[bar width=1.5em, ylabel={\phantomsubcaption \label{fig:UResults:Instability} (b) Instability}]
                    \addplot[fill=fhgBlue,draw=none] coordinates{(8,0) (8,1) (9,2) (8,3)};
                    \addplot[fill=fhgBlue!66,draw=none] coordinates{(4,0) (4,1) (4,2) (6,3)};
                    \addplot[fill=black!20,draw=none] coordinates{(4,0) (3,1) (3,2) (1,3)};
                    \addplot[fill=fhgOrange!66,draw=none] coordinates{(1,0) (3,1) (1,2) (3,3)};
                    \addplot[fill=fhgOrange,draw=none] coordinates{(1,0) (0,1) (1,2) (0,3)};
                \nextgroupplot[bar width=1.5em]
                    \addplot[fill=fhgBlue,draw=none] coordinates{(7,0) (6,1) (6,2) (5,3)};
                    \addplot[fill=fhgBlue!66,draw=none] coordinates{(4,0) (8,1) (8,2) (5,3)};
                    \addplot[fill=black!20,draw=none] coordinates{(2,0) (2,1) (2,2) (2,3)};
                    \addplot[fill=fhgOrange!66,draw=none] coordinates{(4,0) (1,1) (2,2) (2,3)};
                    \addplot[fill=fhgOrange,draw=none] coordinates{(1,0) (1,1) (0,2) (4,3)};
                \nextgroupplot[bar width=1.5em]
                    \addplot[fill=fhgBlue,draw=none] coordinates{(7,0) (10,1) (13,2) (8,3)};
                    \addplot[fill=fhgBlue!66,draw=none] coordinates{(5,0) (4,1) (3,2) (5,3)};
                    \addplot[fill=black!20,draw=none] coordinates{(2,0) (2,1) (1,2) (3,3)};
                    \addplot[fill=fhgOrange!66,draw=none] coordinates{(2,0) (1,1) (0,2) (1,3)};
                    \addplot[fill=fhgOrange,draw=none] coordinates{(1,0) (0,1) (0,2) (0,3)};
                \nextgroupplot[bar width=1.5em, ylabel={\phantomsubcaption \label{fig:UResults:Quality} (c) Quality}]
                    \addplot[fill=fhgBlue,draw=none] coordinates{(0,0) (1,1) (1,2) (0,3)};
                    \addplot[fill=fhgBlue!66,draw=none] coordinates{(2,0) (3,1) (5,2) (4,3)};
                    \addplot[fill=black!20,draw=none] coordinates{(7,0) (8,1) (10,2) (3,3)};
                    \addplot[fill=fhgOrange!66,draw=none] coordinates{(6,0) (5,1) (2,2) (9,3)};
                    \addplot[fill=fhgOrange,draw=none] coordinates{(3,0) (1,1) (0,2) (2,3)};
                \nextgroupplot[bar width=1.5em]
                    \addplot[fill=fhgBlue,draw=none] coordinates{(0,0) (0,1) (0,2) (6,3)};
                    \addplot[fill=fhgBlue!66,draw=none] coordinates{(2,0) (3,1) (3,2) (2,3)};
                    \addplot[fill=black!20,draw=none] coordinates{(9,0) (6,1) (9,2) (7,3)};
                    \addplot[fill=fhgOrange!66,draw=none] coordinates{(3,0) (8,1) (5,2) (2,3)};
                    \addplot[fill=fhgOrange,draw=none] coordinates{(4,0) (1,1) (1,2) (1,3)};
                \nextgroupplot[bar width=1.5em]
                    \addplot[fill=fhgBlue,draw=none] coordinates{(0,0) (0,1) (0,2) (0,3)};
                    \addplot[fill=fhgBlue!66,draw=none] coordinates{(1,0) (0,1) (0,2) (2,3)};
                    \addplot[fill=black!20,draw=none] coordinates{(6,0) (5,1) (6,2) (8,3)};
                    \addplot[fill=fhgOrange!66,draw=none] coordinates{(7,0) (7,1) (7,2) (3,3)};
                    \addplot[fill=fhgOrange,draw=none] coordinates{(3,0) (5,1) (4,2) (4,3)};
            \end{groupplot}
        \end{tikzpicture}
    \end{subfigure}
    \begin{subfigure}{\linewidth}
        \centering
        \begin{tikzpicture}
            \matrix[draw, matrix of nodes, anchor=north, column sep=2pt, row sep=0pt] at (0,0) {
                \node {\colorbox{fhgBlue}{\phantom{x}}}; &
                \node {\colorbox{fhgBlue!66}{\phantom{x}}}; &
                \node {\colorbox{black!20}{\phantom{x}}}; &
                \node {\colorbox{fhgOrange!66}{\phantom{x}}}; &
                \node {\colorbox{fhgOrange}{\phantom{x}}};\\
                \node[align=center, anchor=north, font=\scriptsize] {1 - Very Little}; &
                \node[align=center, anchor=north, font=\scriptsize] {2 - Little}; &
                \node[align=center, anchor=north, font=\scriptsize] {3 - Acceptable}; &
                \node[align=center, anchor=north, font=\scriptsize] {4 - Much}; &
                \node[align=center, anchor=north, font=\scriptsize] {5 - Very Much};\\
            };
        \end{tikzpicture}
    \end{subfigure}
    \caption{User responses in the three scenarios. For Cuttlefish (CF) and Dynamic Tree Colors with three different stability ratios.}
    \label{fig:UResults}
\end{figure}
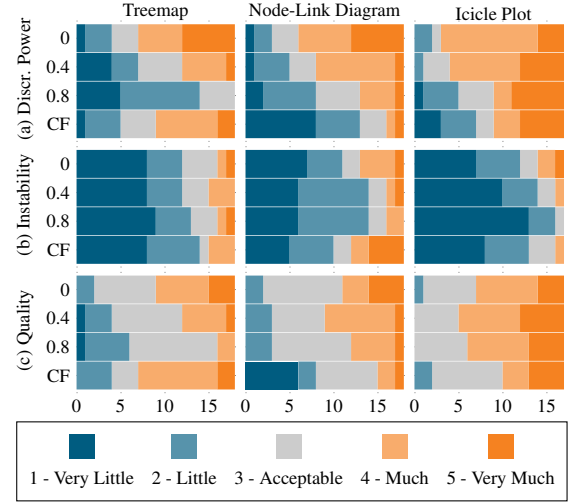

\Cref{fig:UResults:DiscriminativePower} shows the responses regarding perceived discriminative power for the three stability ratios in all three scenarios.
\textbf{H2} states that there should be a decreasing trend in the distributions as the stability ratio increases.
This trend is clearly visible in the treemap and node-link diagram scenarios, but not in the icicle plot scenario, where the number of overall positive responses (values 4 \& 5) decreased while, counter to our hypothesis, the number of very positive (value 5) responses increased.

\Cref{fig:UResults:Instability} shows the perceived confusion due to color instability for the three stability ratios in all three scenarios.
According to \textbf{H1}, we expect a decreasing trend here as well.
A slight such trend can be observed in the node-link diagram and the icicle plot scenarios, but not in the treemap scenario, where all three configurations perform almost identically.

The fact that we observed different trends among the analysis scenarios when investigating the prior two hypotheses already suggests that the scenarios differ with respect to their color map requirements.
But to investigate these differences directly (\textbf{H3}), we introduce a merged scale for color map quality (see \cref{fig:UResults:Quality}).
We compute this merged score by inverting the scale for instability and computing the average between both ratings for each trial (rounded down).
For the treemap scenario, the best quality is achieved with a stability ratio of $0$, while the best is $0.4$ for the icicle plot.
For the node-link diagram, the optimum is not quite clear.
The stability ratio $0.4$ yielded the most positive responses (values 4 or 5), but the number of very high ratings (value 5) is larger for stability ratio $0$.
Regardless, the results confirm that the stability ratio optimum depends on the scenario.

For \textbf{H4}, we compare the quality of \cuttle{ }with the best stability ratio in each scenario.
In the treemap scenario, the quality of \cuttle{ }received more positive responses than any \dtc{ }configuration.
For the other two scenarios, the better-performing stability ratios received more positive responses than \cuttle.
This indicates, that \dtc{ }outperforms \cuttle{ }in the icicle plot and node-link diagram scenarios, but \cuttle{ }is better in the treemap scenario.

During the work on the tasks, participants notably had much less trouble with the color changes than we expected.
This is also reflected in the overall low ratings for the perceived confusion due to the color instability (\cref{fig:UResults:Instability}).
This is partially explained by some of the participants' remarks, one stating that \quot{I have a feeling that one does not even try to remember the colors, because they are changing all the time.} and another that \quot{The colors didn't interest me, I went with the position instead}.
This indicates that some participants avoided memorizing colors as identifying attribute, because they expected them to change.
In the brief trials of our study, relying on other visual channels enabled participants to complete the tasks without much confusion.
The application of such strategies has impacted our measurements of perceived color instability.
This effect also varied throughout the analysis scenarios (\quot{I think that the colors are much more usable for the icicle plot than especially for the treemap}).
This provides partial explanation why we observed the expected trend most strongly in the icicle plot scenario.
Designs of future studies into the effects of color stability should make sure to explicitly take such strategies into account.
Nevertheless, participants were able to assess discriminative power throughout the trials (\quot{I can distinguish the Euro vs. not-Euro [countries], but within each [group], they are all the same.}), which is reaffirmed by the much stronger trends in those responses.

\section{\uppercase{Discussion}}

Both the results of the quality analysis and the user study (\textbf{H1} \& \textbf{H2}) indicate that \dtc{ }achieves its purpose, by allowing a flexible adjustment of the tradeoff between discriminative power and color stability.
However, we were not able to observe the expected trend in the discriminative power within the icicle plot scenario as well as within the color instability in the treemap scenario.
Our investigation of \textbf{H3} indicates that the relationship between stability ratio and user perception depends on the analysis scenario.
Thus, it is possible that in some scenarios, this relationship is not of a linear nature.
But further research is needed to investigate these deviations from our expectations.

We were also able to show that our algorithm is more flexible than \cuttle, outperforming the algorithm both in terms of measured quality as well as user perception (\textbf{H4}) in most application scenarios.
While \dtc{ }manages to achieve a lower reference color instability than \cuttle, it unexpectedly also achieves a lower incremental color instability.
However, when specifically analyzing the treemap scenario, both the quality metrics as well as the user study results show a superior performance for \cuttle.
This indicates that \dtc{ }is the more flexible algorithm, that is applicable to many application scenarios while delivering good results.
But the results also show that an algorithm specifically tuned for a given application scenario can outperform \dtc.
Hence, for the multiscale visualization scenario, which only displays one hierarchy level's colors at a time and allows a zoom into individual hierarchy branches, we recommend to apply \cuttle{ }instead of \dtc.

As reported above, participants were able to develop strategies to avoid relying on color during the experiment.
After finding one such strategy for a given scenario, color changes were no longer a confusing factor, which explains the low ratings for the perceived impact of color instability.
However, the fact that participants likely have discovered certain strategies at different times throughout the study induces a measuring inaccuracy in our results, reaffirming our decision to proceed with a solely qualitative interpretation.
In addition, the short trial duration for individual tasks in our study did likely not facilitate the construction of a mental map of the color map.
Hence, we could not measure the impact of color instability regarding mental map preservation.
These open questions should be addressed in future user studies of dynamic color maps.

\section{\uppercase{Conclusion}}

In this paper, we have presented the \dtc{ }algorithm for the dynamic generation of hierarchical color maps.
We have defined quantitative metrics for discriminative power and color instability, to facilitate the empirical discussion of hierarchical color map quality.
Furthermore, we investigated the results of \dtc{ }in comparison with \cuttle, in terms of our defined quality criteria as well as a qualitative user study.
Our findings overall confirm that \dtc{ }is the more flexible algorithm that can be applied to many analysis scenarios while delivering good results, yet it can not reach \cuttle's performance in the specific scenario that it was designed for.
However, since the approach of \dtc{ }only relies on a rigid rotation and an interpolation scheme, it is compatible with other static color map algorithms, thus leaving much room for future improvements.
While our studies provide preliminary insight into the impact of the tradeoff between the two metrics, our quality analysis should be extended with a definition of color map uniformity and the user study's design should be improved, based on our learnings, to yield conclusive evidence.

\section*{\uppercase{Supplemental Materials}}

All supplemental materials are available on OSF at \url{https://osf.io/xfp8n/}, released under a CC-BY 4.0 license. In particular, they include (1) additional screenshots of \dtc{ }results, including the altered application examples as they were used in the user study, (2) the code-structure datasets, (3) the results of the quality analysis and the user study.

\section*{\uppercase{Acknowledgements}}

This research work has been funded by the German Federal Ministry of Education and Research and the Hessian Ministry of Higher Education, Research, Science and the Arts within their joint support of the National Research Center for Applied Cybersecurity ATHENE.

\bibliographystyle{apalike}
{\small
\bibliography{ref.bib}}

\end{document}